\documentclass[preprint,12pt]{elsarticle}
\usepackage{amssymb,a4wide,latexsym,makeidx,xy,epsfig,fleqn,amsmath,amsfonts,amscd,extpfeil}
\usepackage{mathrsfs}
\usepackage{booktabs}
\usepackage{lineno} 
\usepackage{multirow}
\usepackage{array}
\usepackage{float}
\usepackage{natbib}
\usepackage{doi}
\usepackage{color}
\usepackage[linesnumbered, ruled]{algorithm2e}
\SetKwRepeat{Do}{do}{while}
\usepackage{graphicx}
\usepackage{subcaption}
\usepackage{enumerate}
\usepackage{booktabs}
\usepackage{amsthm}  
\usepackage{amsmath}
\usepackage{amssymb}
\usepackage{bm}
\usepackage{graphicx}
\usepackage{caption}
\usepackage{natbib}
\usepackage{cite}
\allowdisplaybreaks[3]
\newtheorem{theorem}{Theorem}
\newtheorem{corollary}{Corollary}

\begin{document}
	\begin{frontmatter}
		
		\title{Optimal bias of utility function between two-layer network for the evolution of prosocial behavior in two-order game and higher-order game}
		
		\author[label1]{Yihe Ma}\ead{yihema@mail.nwpu.edu.cn}

		\address[label1]{School of Mathematics and Statistics, Northwestern
			Polytechnical University, \\ Xi'an,  710129,  China}

		\begin{abstract}
			
			Cooperation is an important research object in economics, sociology, and biology, and the evolution of cooperation in structured populations is a interesting research topic. We mainly focus on the evolution of cooperation with two-order and higher-order game in two-layer network. We introduce a bias coefficient of utility function and study the influence of bias coefficient on the evolution of cooperation in two-layer network. We firstly provide theoretical analysis of fixation probabilities of two-order and higher-order game under weak selection in two-layer network.Secondly,based on the expression of fixation probability, we obtain the critical value of the two different games by comparing the size relationship of fixation probability under weak selection condition and neutral selection condition. Finally, by comparing the relationship between the critical value of single-layer and two-layer network in two-order game and higher-order game, when the nonlinear factor satisfies certain conditions, it is concluded that when the optimal bias coefficient tends towards 0 is met, some two-layer networks promote the evolution of cooperative behavior more than some single-layer networks. 
			
		\end{abstract}
	\end{frontmatter}

	\section{Introduction}
	Cooperation is an important research object in fields such as sociology and biology, and the evolutionary of cooperation have become a hot research topic\citep{allen2023symmetry}\citep{allen2024nonlinear}\citep{mcavoy2022evaluating}\citep{sheng2023evolutionary}\citep{wang2023high}\citep{mcavoy2022evaluating}\citep{zhong2022dynamical}. There have been important theoretical results on the evolution of cooperative behavior in finite populations and structured finite populations\citep{ohtsuki2006simple}\citep{traulsen2006stochasticity}\citep{lieberman2005evolutionary}\citep{taylor2004evolutionary}\citep{hauert2004spatial}\citep{nowak2004emergence}\citep{broom2022game}. Some researchers have obtained the evolution laws of cooperation in single-layer and two-layer networks under two-order game, and pointed out that the evolution of cooperative behavior is more promoted in two-layer network\citep{su2022evolution}. In addition, similar conclusions have also been drawn in the experimental simulation analysis of single-layer and two-layer networks. We know that higher-order games in single-layer network promote the evolution of cooperation more than two-order games\citep{sheng2024strategy}. Therefore, we also focus on higher-order game on two-layer network.
	
	The main focus of this article is on how the optimal bias coefficient of utility function $\alpha$ for two-order game in two-layer network affects the evolution of cooperation.We could have the fixation probability of two-order game and higher-order game under weak selection. By calculating the $(\frac{b}{c})^{*}$ with fixation probability, it can be concluded that the evolution of cooperative behavior in some two-layer network is superior to that in some single-layer network. We can conclude that  when the bias coefficient $\alpha$ tends towards 0, some two-layer networks structures with two-order game and higher-order game will most promote cooperation, when the nonlinear factor satisfies certain conditions.
	\section{Model}
	Our model is a finite population with an interdependent two-layer network structure. Each layer of an interdependent two-layer network is represented as $[1]$ and $[2]$, respectively. There are $N_{[1]}$ individuals in the $[1]$ network and there are $N_{[2]}$ individuals in the $[2]$ network.
	
	We focus on the evolutionary dynamics of two behavioral strategies, $C$(cooperator) and $D$(defector), in a population.  The state of population is represented\citep{mcavoy2021fixation} by $\textbf{s}\in{{{\{0,1\} }^{N_{[1]}}}\times{{\{0,1\} }^{N_{[2]}}}}$, where $s_{i_{[l]}}=1$(resp.$s_{i_{[l]}}=0$) indicates behavioral strategy of the $i$ individual in $l$ network is $C$(resp.$D$). In the following text, we use the symbol $[l],l\in{\{0,1\}}$ to represent individuals in the $[l]$ layer network or values related to the $l_{[1]}$($l_{[2]}$) network. We let ${\textbf{C}}_{[1]}$ indicate the state that all individuals have type $C$ in $l_{(1)}$ network, ${{\textbf{C}}_{(1)}} \times {{\textbf{C}}_{(2)}}$ denote the state that all individuals in the two-layer network are assigned type $C$, the definition of sign ${\textbf{C}}_{(2)}$, ${\textbf{D}}_{(1)}$, ${\textbf{D}}_{(2)}$, ${{\textbf{C}}_{(1)}} \times {{\textbf{D}}_{(2)}}$, ${{\textbf{D}}_{(1)}} \times {{\textbf{C}}_{(2)}}$, ${{\textbf{D}}_{(1)}} \times {{\textbf{D}}_{(2)}}$ are similar. For brevity, we write $\mathbb{A}:={\{0,1}\}$, ${\mathbb{A}}_{T}^{N_{(1)}}:={\mathbb{A}}^{N_{(1)}} - \{{\textbf{C}}_{(1)}, {\textbf{D}}_{(1)}\}$, ${\mathbb{A}}_{T}^{N_{(2)}}:={\mathbb{A}}^{N_{(2)}} - \{{\textbf{C}}_{(2)}, {\textbf{D}}_{(2)}\}$, then the state of population is represented by $\textbf{s}\in{{\mathbb{A}}^{N_{(1)}}} \times {{\mathbb{A}}^{N_{(2)}}}$\citep{mcavoy2021fixation}.
	
	We consider the weighted, undirected interdependent two-layer network structure, the $l_{(1)}$ network with adjacency matrix\citep{su2022evolution} $(w_{i_{(1)}j_{(1)}})_{i_{(1)},j_{(1)}=1}^{N_{(1)}}$, the $l_{(2)}$ network with adjacency matrix $(w_{i_{(2)}j_{(2)}})_{i_{(2)},j_{(2)}=1}^{N_{(2)}}$, the undirected two-layer network can be represented by $w_{i_{(1)}j_{(1)}}=w_{j_{(1)}i_{(1)}}$, $w_{i_{(2)}j_{(2)}}=w_{j_{(2)}i_{(2)}}$. Below we will introduce the connection relationship between three individuals in each layer network. In $l_{(1)}$ network, if individual $i_{(1)}$, $j_{(1)}$, $k_{(1)}$ can form a triangle between each other, then $w_{i_{(1)}j_{(1)}k_{(1)}}=1$, otherwise, $w_{i_{(1)}j_{(1)}k_{(1)}}=0$, the expression of the $l_{(2)}$ network is similar. We denote by $(a_{i_{(1)}j_{(2)}})_{i_{(1)},j_{(2)}=1}^{N_{(1)},N_{(2)}}$ the interdependence strength matrix of individuals in $l_{(1)}$ network to individuals in $l_{(2)}$ network, similarly, the interdependence strength matrix of individuals in $l_{(1)}$ network to individuals in $l_{(2)}$ network with $(b_{i_{(2)}j_{(1)}})_{i_{(2)},j_{(1)}=1}^{N_{(2)},N_{(1)}}$. For concise expression, we adopt $w_{i_{(1)}}:=\sum_{j_{(1)}=1}^{N_{(1)}} {w_{i_{(1)}j_{(1)}}}$ and the one step transition probability is $p_{i_{(1)}j_{(1)}}^{(1)}:=\frac{w_{i_{(1)}j_{(1)}}}{w_{i_{(1)}}}$
	in $l_{(1)}$ network, equally, we let $p_{i_{(2)}j_{(2)}}^{(1)}$ represent the one step transition probability in $l_{(2)}$ network\citep{mcavoy2021fixation}. We define $p_{i_{(1)}j_{(1)}}^{(0)}$ equals $1$ when $i_{(1)}=j_{(1)}$ and other situations are $0$, $p_{i_{(2)}j_{(2)}}^{(0)}$ equals $1$ when $i_{(2)}=j_{(2)}$ and other situations are $0$\citep{mcavoy2021fixation}. We adopt $a_{i_{(1)}}:=\sum_{j_{(2)}=1}^{N_{(2)}} {a_{i_{(1)}j_{(2)}}}$ and the interdependence proportion $q_{i_{(1)}j_{(2)}}:=\frac{a_{i_{(1)}j_{(2)}}}{a_{i_{(1)}}}$
	in $l_{(1)}$ network, equally, we let $q_{i_{(2)}j_{(1)}}$ denote the interdependence proportion in $l_{(2)}$ network.
	
	Within each layer of the network, we firstly focus on a particular game called Prisoner's Dilemma game, which is only the two-order game. The utility function is expressed in the form of equation(1)(2)\citep{mcavoy2021fixation}, where $b > c > 0$. Between two layers of network, we introduce the bias coefficient $\alpha \in [0,1]$ to determine the bias in each layer.When the deviation coefficient is 0, it indicates that the utility of individuals in one layer of the network is completely controlled by the utility of individuals in another layer of the network; when the deviation coefficient is 1, it represents the same utility situation as the individual in the single-layer network. By utilizing a utility feedback mechanism in a two-layer network, we obtain within layer utility for each individual in $l_{(1)}$ network is
	\begin{equation}\label{eq:example}
		{u_{i_{(1)}}} (\textbf{s})=-{w_{i_{(1)}}}c{s_{i_{(1)}}}+\sum_{k_{(1)}=1}^{N_{(1)}} {w_{i_{(1)}k_{(1)}}}b{s_{k_{(1)}}},
	\end{equation}
	and within layer payoff for each individual in $l_{(2)}$ network is
	\begin{equation}\label{eq:example}
		{u_{i_{(2)}}} (\textbf{s})=-{w_{i_{(2)}}}c{s_{i_{(2)}}}+\sum_{k_{(2)}=1}^{N_{(2)}} {w_{i_{(2)}k_{(2)}}}b{s_{k_{(2)}}}.
	\end{equation}
	
	We could obtain the utility of the higher-order game which is a combination of two-order and three-order games in every layer, three-order game refers to a game involving three participants\citep{sheng2024strategy}, taking individuals in the $l_{(1)}$ network as an example.
	\begin{equation}\label{eq:example}
		\begin{aligned}
			&{u_{i_{(1)}}} (\textbf{s})=
			\sum_{k_{(1)}=1}^{N_{(1)}}{w_{i_{(1)}k_{(1)}}}\biggl((\frac{b}{2}-c){s_{i_{(1)}}}+\frac{b}{2}{s_{k_{(1)}}}+b\frac{\theta_{2}-1}{2}{s_{i_{(1)}}}{s_{k_{(1)}}}\biggr) \\
			&\quad\quad\quad\quad+
			\frac{1}{2}\sum_{k_{(1)},j_{(1)}=1}^{N_{(1)}}w_{i_{(1)}k_{(1)}j_{(1)}}\biggl((\frac{b}{3}-c){s_{i_{(1)}}}+\frac{b}{3}({s_{i_{(1)}}}+{s_{k_{(1)}}}+{s_{j_{(1)}}}) \\
			&\quad\quad\quad\quad+
			b\frac{\theta_{3}-1}{3}({s_{i_{(1)}}}{s_{k_{(1)}}}+{s_{i_{(1)}}}{s_{j_{(1)}}}+{s_{k_{(1)}}}{s_{j_{(1)}}})+\frac{{(\theta_{3}-1)}^{2}}{3}{s_{i_{(1)}}}{s_{k_{(1)}}}{s_{j_{(1)}}}\biggr),
		\end{aligned}
	\end{equation}
	where the coefficient $\theta_{2}$ and $\theta_{2}$ are nonlinear factor. We obtain total utility for each individual in $l_{(1)}$ network is\citep{su2022evolution}
	\begin{equation}\label{eq:example}
		\tilde{u}_{i_{(1)}}(\textbf{s})=\alpha{u}_{i_{(1)}}(\textbf{s})+(1-\alpha)\sum_{k_{(2)}=1}^{N_{(2)}} {q_{i_{(1)}k_{(2)}}^{(1)}}{u_{k_{(2)}}}(\textbf{s}),
	\end{equation} 
	and total utility for each individual in $l_{(2)}$ network is
	\begin{equation}\label{eq:example}
		\tilde{u}_{i_{(2)}}(\textbf{s})=\alpha{u}_{i_{(2)}}(\textbf{s})+(1-\alpha)\sum_{k_{(1)}=1}^{N_{(1)}} {q_{i_{(2)}k_{(1)}}^{(1)}}{u_{k_{(1)}}}(\textbf{s}).
	\end{equation} 
	We introduce individual fecundity related to total individual utility in $l_{(1)}$ network, $F_{i_{(1)}}$,
	\begin{equation}\label{eq:example}
		F_{i_{(1)}}:=1+\delta{\tilde{u}_{i_{(1)}}(\textbf{s})},
	\end{equation}
	and individual fecundity in $l_{(2)}$ network, $F_{i_{(2)}}$,
	\begin{equation}\label{eq:example}
		F_{i_{(2)}}:=1+\delta{\tilde{u}_{i_{(2)}}(\textbf{s})},
	\end{equation}
	where $\delta \geq 0$ is the selection intensity parameter,${\delta}=0$ represents neutral drift and $\delta \ll 1$ represents weak selection, we let superscript $\circ$ denote the value under neutral drift\citep{mcavoy2021fixation}.
	
	Within each layer of the network, we consider classic death-Birth updating\citep{mcavoy2021fixation}, With this update rule, the probability of individual $j_{(1)}$ imitating individual $i_{(1)}$ strategy in $l_{(1)}$ network is
	\begin{equation}\label{eq:example}
		e_{i_{(1)}j_{(1)}}(\textbf{s})={\frac{1}{N_{(1)}}}\frac{{F_{i_{(1)}}}(\textbf{s}){w_{i_{(1)}j_{(1)}}}}{\sum_{l_{(1)}=1}^{N_{(1)}} {F_{l_{(1)}}}(\textbf{s}){w_{l_{(1)}j_{(1)}}}},		
	\end{equation}
	and the probability of individual $j_{(2)}$ imitating individual $i_{(1)}$ strategy in $l_{(2)}$ network is
	\begin{equation}\label{eq:example}
		e_{i_{(2)}j_{(2)}}(\textbf{s})={\frac{1}{N_{(2)}}}\frac{{F_{i_{(2)}}}(\textbf{s}){w_{i_{(2)}j_{(2)}}}}{\sum_{l_{(2)}=1}^{N_{(2)}} {F_{l_{(2)}}}(\textbf{s}){w_{l_{(2)}j_{(2)}}}}.		
	\end{equation}
	\section{Result}
	Without artificial "mutation",the population eventually reach a monomorphic state in which individuals on the same layer have the same state\citep{mcavoy2021fixation}. We let fixation probability,$\rho_{C_{(1)} \times C_{(2)}}(\bm{\eta})$, which denote the probability of transitioning from the initial state $\bm{\eta}$,$\bm{\eta} \in {{\mathbb{A}}_{T}^{N_{(1)}} \times {\mathbb{A}}_{T}^{N_{(2)}}}$ to the state $C_{(1)} \times C_{(2)}$ in the end\citep{mcavoy2021fixation}. In order to generalize the conclusion, we calculate fixation probability of weak selection under random initial states by introducing mutant-appearance distributions $\lambda$\citep{mcavoy2021fixation}. According to the condition $E_{\lambda}[\rho_{C_{(1)} \times C_{(2)}}(\bm{\eta})+\rho_{{\textbf{C}}_{(1)} \times {\textbf{D}}_{(2)}}(\bm{\eta})] > E_{\lambda}^{\circ}[\rho_{C_{(1)} \times C_{(2)}}^{\circ}(\bm{\eta})+\rho_{C_{(1)} \times D_{(2)}}^{\circ}(\bm{\eta})]$, we can explain whether population structure promotes cooperative behavior to replace betrayal behavior in the $l_{(1)}$ network, the $l_{(2)}$ network is similar \citep{imhof2006evolutionary}\citep{su2022evolution}\citep{mcavoy2021fixation}.
	
	To calculate $E_{\lambda}[\rho_{C_{(1)} \times C_{(2)}}(\bm{\eta})+\rho_{C_{(1)} \times D_{(2)}}(\bm{\eta})]$ and $E_{\lambda}[\rho_{C_{(1)} \times C_{(2)}}(\bm{\eta})+\rho_{D_{(1)} \times C_{(2)}}(\bm{\eta})]$, we need to introduce a few concepts.Primarily, we give the definition of reproductive value(RV) of individual $i_{(1)}$, represented by symbol $v_{i_{(1)}}$, it is the only solution to the following equation\citep{mcavoy2021fixation}
	\begin{equation}\label{eq:example}
		\sum_{j_{(1)}=1}^{N_{(1)}} e_{i_{(1)}j_{(1)}}^{\circ}v_{j_{(1)}} = \sum_{j_{(1)}=1}^{N_{(1)}} e_{j_{(1)}i_{(1)}}^{\circ}v_{i_{(1)}},
	\end{equation}
	\begin{equation}\label{eq:example}
		\sum_{i_{(1)}=1}^{N_{(1)}} v_{i_{(1)}} =1.
	\end{equation}
	the reproductive value(RV) of individual $i_{(2)}$, $v_{i_{(2)}}$, it is the only solution to the following equation  
	\begin{equation}\label{eq:example}
		\sum_{j_{(2)}=1}^{N_{(2)}} e_{i_{(2)}j_{(2)}}^{\circ}v_{j_{(2)}} = \sum_{j_{(2)}=1}^{N_{(2)}} e_{j_{(2)}i_{(2)}}^{\circ}v_{i_{(2)}},
	\end{equation}
	\begin{equation}\label{eq:example}
		\sum_{i_{(2)}=1}^{N_{(2)}} v_{i_{(2)}} =1.
	\end{equation}
	
	Next, for any state $\textbf{s}$, the RV-weighted frequency in two-layer network can be defined as $\hat{s}_{(1)}:={\sum_{i_{(1)}=1}^{N_{(1)}} v_{i_{(1)}}s_{i_{(1)}}}$, $\hat{s}_{(2)}:= {\sum_{i_{(2)}=1}^{N_{(2)}} v_{i_{(2)}}s_{i_{(2)}}}$\citep{mcavoy2021fixation}.
	We note subsets $\textbf{I}\subseteq \{1_{(1)},...,N_{(1)},1_{(2)},...,N_{(2)}\}$, and $\textbf{s}_{\textbf{I}}:=\prod_{i\in \textbf{I}} s_{i}$; $\textbf{I}_{(1)}\subseteq \{1_{(1)},...,N_{(1)}\}$, and $\textbf{s}_{\textbf{I}_{(1)}}:=\prod_{i_{(1)}\in \textbf{I}_{(1)}} s_{i_{(1)}}$;$\textbf{I}_{(2)}\subseteq \{1_{(2)},...,N_{(2)}\}$, and $\textbf{s}_{\textbf{I}_{(2)}}:=\prod_{i_{(2)}\in \textbf{I}_{(2)}} s_{i_{(2)}}$. To simplify notation, we use $\textbf{s}_{i_{(1)}j_{(1)}}$ instead of $\textbf{s}_{\{i_{(1)},j_{(1)}\}}$. Lastly,we introduce two expression types of mutant-appearance distribution $\lambda$\citep{mcavoy2021fixation},where uniform initialization is
	\begin{equation}\label{eq:example}
		\lambda({\textbf{1}_{i_{(1)}}^{(1)}} \times {\textbf{1}_{i_{(2)}}^{(2)}}) = \frac{1}{N_{(1)} \times N_{(2)}},
	\end{equation}
	the ordinary initialization is\citep{mcavoy2021fixation} 
	\begin{equation}\label{eq:example}
		\begin{aligned}
			\lambda({\textbf{1}_{i_{(1)}}^{(1)}} \times {\textbf{1}_{i_{(2)}}^{(2)}}) =
			\lambda_{D}({\overline{{\textbf{1}_{i_{(1)}}^{(1)}} \times {\textbf{1}_{i_{(2)}}^{(2)}}}}) = v_{i_{(1)}i_{(2)}},
		\end{aligned}
	\end{equation}
	where $\overline{\eta_{i}}=1-\eta_{i}$,$v_{i_{(1)}i_{(2)}}$ represents probability,
	where temperature initialization is
	\begin{equation}\label{eq:example}
		\lambda({\textbf{1}_{i_{(1)}}^{(1)}} \times {\textbf{1}_{i_{(2)}}^{(2)}}) = \frac{d_{i_{(1)}}({{\textbf{D}}_{(1)}} \times {{\textbf{D}}_{(2)}}) + d_{i_{(2)}}({{\textbf{D}}_{(1)}} \times {{\textbf{D}}_{(2)}})}{{\sum_{j_{(1)}=1}^{N_{(1)}} d_{j_{(1)}}({{\textbf{D}}_{(1)}} \times {{\textbf{D}}_{(2)}})} + {\sum_{j_{(2)}=1}^{N_{(2)}} d_{j_{(2)}}({{\textbf{D}}_{(1)}} \times {{\textbf{D}}_{(2)}})}}.
	\end{equation}
	where $d_{i_{(1)}}(\textbf{s}):=\sum_{j_{(1)}=1}^{N_{(1)}} e_{j_{(1)}i_{(1)}}(\textbf{s})$ denote the death rate of $i_{(1)}$ in $l_{(1)}$ network, similarly, we use $d_{i_{(2)}}(\textbf{s})$ to represent the death rate of $i_{(2)}$ in $l_{(2)}$ network.
	
	Through a large amount of complex calculations, we can draw the following conclusion of two-order game in each layer network under weak selection\citep{su2022evolution}\citep{mcavoy2021fixation}, when state $\bm{\eta}$ satisfies condition $\bm{\eta} \in {{\mathbb{A}}_{T}^{N_{(1)}} \times {\mathbb{A}}_{T}^{N_{(2)}}}$
	\begin{equation}\label{eq:example}
		\begin{aligned}
			&E_{\lambda}[\rho_{{\textbf{C}}_{(1)} \times {\textbf{C}}_{(2)}}(\bm{\eta})+\rho_{{\textbf{C}}_{(1)} \times {\textbf{D}}_{(2)}}(\bm{\eta})]=E_{\lambda}[\hat{\eta}_{(1)}] \\
			&\quad\quad\quad\quad+
			\frac{\delta}{N_{(1)}}\biggl(\sum_{i_{(1)}=1}^{N_{(1)}} v_{i_{(1)}}\sum_{j_{(1)}=1}^{N_{(1)}} \sum_{k_{(1)}=1}^{N_{(1)}} M_{k_{(1)}}^{i_{(1)}j_{(1)}} \left( \psi_{i_{(1)}k_{(1)}}^{\lambda} -\psi_{j_{(1)}k_{(1)}}^{\lambda} \right)  \\
			&\quad\quad\quad\quad+
			\sum_{i_{(1)}=1}^{N_{(1)}} v_{i_{(1)}}\sum_{j_{(1)}=1}^{N_{(1)}} \sum_{k_{(2)}=1}^{N_{(2)}} M_{k_{(2)}}^{i_{(1)}j_{(1)}}\left( \phi_{i_{(1)}k_{(2)}}^{\lambda} -\phi_{j_{(1)}k_{(2)}}^{\lambda} \right) \biggr) \\
			&\quad\quad\quad\quad+
			\mathcal{O}(\delta^2),
		\end{aligned}
	\end{equation}
	where 
	\begin{equation}\label{eq:example}
		\psi_{i_{(1)}j_{(1)}}^{\lambda} = \frac{N_{(1)}}{2}E_{\lambda}^{\circ}[\hat{\eta}_{(1)} - \eta_{i_{(1)}j_{(1)}}] + \frac{1}{2}\sum_{k_{(1)}=1}^{N_{(1)}} \left(p_{i_{(1)}k_{(1)}}^{(1)}\psi_{k_{(1)}j_{(1)}}^{\lambda} + p_{j_{(1)}k_{(1)}}^{(1)}\psi_{i_{(1)}k_{(1)}}^{\lambda}\right),
	\end{equation}
	\begin{equation}\label{eq:example}
		\begin{aligned}
			&\phi_{i_{(1)}j_{(2)}}^{\lambda} = E_{\lambda}^{\circ}[\frac{N_{(1)}N_{(2)}}{N_{(1)}+N_{(2)}-1}((\hat{\eta}_{(1)}\hat{\eta}_{(2)}) -\eta_{i_{(1)}j_{(2)}})] \\
			&\quad\quad\quad\quad+
			\frac{1}{N_{(1)}+N_{(2)}-1}\left(\sum_{k_{(1)}=1}^{N_{(1)}}\sum_{k_{(2)}=1}^{N_{(2)}} p_{i_{(1)}k_{(1)}}^{(1)} p_{j_{(2)}k_{(2)}}^{(1)}\phi_{k_{(1)}k_{(2)}}^{\lambda}\right) \\
			&\quad\quad\quad\quad+
			\frac{N_{(2)}-1}{N_{(1)}+N_{(2)}-1}\sum_{k_{(1)}=1}^{N_{(1)}}p_{i_{(1)}k_{(1)}}^{(1)}\phi_{k_{(1)}j_{(2)}}^{\lambda}+\frac{(N_{(1)}-1)N_{(1)}}{(N_{(1)}+N_{(2)}-1)N_{(2)}}\sum_{k_{(2)}=1}^{N_{(2)}}p_{j_{(2)}k_{(2)}}^{(1)}\phi_{i_{(1)}k_{(2)}}^{\lambda}, \\
		\end{aligned}
	\end{equation}
	\begin{equation}\label{eq:example}
		\begin{aligned}
			&E_{\lambda}[\rho_{{\textbf{C}}_{(1)} \times {\textbf{C}}_{(2)}}(\bm{\eta})+\rho_{{\textbf{D}}_{(1)} \times {\textbf{C}}_{(2)}}(\bm{\eta})]=E_{\lambda}[\hat{\eta}_{2}] \\
			&\quad\quad\quad\quad+
			\frac{\delta}{N_{(2)}}\biggl(\sum_{i_{(2)}=1}^{N_{(2)}} v_{i_{(2)}}\sum_{j_{(2)}=1}^{N_{(2)}} \sum_{k_{(2)}=1}^{N_{(2)}} M_{k_{(2)}}^{i_{(2)}j_{(2)}}\left( \xi_{i_{(2)}k_{(2)}}^{\lambda} -\xi_{j_{(2)}k_{(2)}}^{\lambda} \right) \\
			&\quad\quad\quad\quad+
			\sum_{i_{(2)}=1}^{N_{(2)}} v_{i_{(2)}}\sum_{j_{(2)}=1}^{N_{(2)}} \sum_{k_{(1)}=1}^{N_{(1)}} M_{k_{(1)}}^{i_{(2)}j_{(2)}}\left( \zeta_{i_{(2)}k_{(1)}}^{\lambda} -\zeta_{j_{(2)}k_{(1)}}^{\lambda} \right)  \biggr)\\
			&\quad\quad\quad\quad+
			\mathcal{O}(\delta^2),
		\end{aligned}
	\end{equation}
	where 
	\begin{equation}\label{eq:example}
		\xi_{i_{(2)}j_{(2)}}^{\lambda} = \frac{N_{(2)}}{2}E_{\lambda}^{\circ}[\hat{\eta}_{(2)} - \eta_{i_{(2)}j_{(2)}}] + \frac{1}{2}\sum_{k_{(2)}=1}^{N_{(2)}} \left(p_{i_{(2)}k_{(2)}}^{(1)}\xi_{k_{(2)}j_{(2)}}^{\lambda} + p_{j_{(2)}k_{(2)}}^{(1)}\xi_{i_{(2)}k_{(2)}}^{\lambda}\right),
	\end{equation}
	\begin{equation}\label{eq:example}
		\begin{aligned}
			&\zeta_{i_{(2)}j_{(1)}}^{\lambda} = E_{\lambda}^{\circ}[\frac{N_{(1)}N_{(2)}}{N_{(1)}+N_{(2)}-1}((\hat{\eta}_{(1)}\hat{\eta}_{(2)}) -\eta_{i_{(2)}j_{(1)}})] \\
			&\quad\quad\quad\quad+
			\frac{1}{N_{(1)}+N_{(2)}-1}\left(\sum_{k_{(1)}=1}^{N_{(1)}}\sum_{k_{(2)}=1}^{N_{(2)}} p_{j_{(1)}k_{(1)}}^{(1)} p_{i_{(2)}k_{(2)}}^{(1)} \zeta_{k_{(2)}k_{(1)}}^{\lambda}\right) \\
			&\quad\quad\quad\quad+
			\frac{N_{(2)}-1}{N_{(1)}+N_{(2)}-1}\sum_{k_{(1)}=1}^{N_{(1)}}p_{j_{(1)}k_{(1)}}^{(1)}\zeta_{k_{(1)}i_{(2)}}^{\lambda}+\frac{N_{(1)}-1}{N_{(1)}+N_{(2)}-1}\sum_{k_{(2)}=1}^{N_{(2)}}p_{i_{(2)}k_{(2)}}^{(1)}\zeta_{j_{(1)}k_{(2)}}^{\lambda}, \\
		\end{aligned}
	\end{equation}
	The above expressions for $\xi_{i_{(2)}j_{(2)}}^{\lambda}, \zeta_{i_{(2)}j_{(1)}}^{\lambda}, \phi_{i_{(1)}j_{(2)}}^{\lambda},\psi_{i_{(1)}j_{(1)}}^{\lambda}$ are only satisfied when the variables such as $i_{(1)},j_{(1)}$ are not equal to each other. Due to the complexity of equations, when there is an equal relationship between variables such as $i_{(1)},j_{(1)}$, the corresponding expressions can be easily obtained, so they will not be repeated.
	
	we also obtain the conclusion of higher-order game in each layer network\citep{su2022evolution}\citep{mcavoy2021fixation}, here we present the conclusion of the $l_{(1)}$ network,
	\begin{equation}\label{eq:example}
		\begin{aligned}
			&\rho_{{\textbf{C}}_{(1)} \times {\textbf{C}}_{(2)}}(\bm{\eta})+\rho_{{\textbf{C}}_{(1)} \times {\textbf{D}}_{(2)}}(\bm{\eta}) =
			\hat{\eta}_{(1)}+\frac{\delta}{N_{(1)}} \biggr(\sum_{i_{(1)}=1}^{N_{(1)}} v_{i_{(1)}}\sum_{j_{(1)}=1}^{N_{(1)}} \sum_{k_{(1)}=1}^{N_{(1)}} M_{k_{(1)}}^{i_{(1)}j_{(1)}} \biggr( \psi_{i_{(1)}k_{(1)}} -\psi_{j_{(1)}k_{(1)}} \biggr)  \\
			&\quad\quad\quad\quad+
			\sum_{i_{(1)}=1}^{N_{(1)}} v_{i_{(1)}}\sum_{j_{(1)}=1}^{N_{(1)}} \sum_{k_{(1)},m_{(1)}=1}^{N_{(1)}} M_{k_{(1)}m_{(1)}}^{i_{(1)}j_{(1)}} \biggr( \nu_{i_{(1)}k_{(1)}m_{(1)}} -\nu_{j_{(1)}k_{(1)}m_{(1)}} \biggr) \\
			&\quad\quad\quad\quad+
			\sum_{i_{(1)}=1}^{N_{(1)}} v_{i_{(1)}}\sum_{j_{(1)}=1}^{N_{(1)}} \sum_{k_{(1)},m_{(1)},n_{(1)}=1}^{N_{(1)}} M_{k_{(1)}m_{(1)}n_{(1)}}^{i_{(1)}j_{(1)}} \biggr( \kappa_{i_{(1)}k_{(1)}m_{(1)}n_{(1)}} -\kappa_{j_{(1)}k_{(1)}m_{(1)}n_{(1)}}\biggr) \\
			&\quad\quad\quad\quad+
			\sum_{i_{(1)}=1}^{N_{(1)}} v_{i_{(1)}}\sum_{j_{(1)}=1}^{N_{(1)}} \sum_{k_{(2)}=1}^{N_{(2)}} M_{k_{(2)}}^{i_{(1)}j_{(1)}}\biggr( \phi_{i_{(1)}k_{(2)}} -\phi_{j_{(1)}k_{(2)}} \biggr) \\
			&\quad\quad\quad\quad+
			\sum_{i_{(1)}=1}^{N_{(1)}} v_{i_{(1)}}\sum_{j_{(1)}=1}^{N_{(1)}} \sum_{k_{(2)},m_{(2)}=1}^{N_{(2)}} M_{k_{(2)}m_{(2)}}^{i_{(1)}j_{(1)}}\biggr( \sigma_{i_{(1)}k_{(2)}m_{(2)}} -\sigma_{j_{(1)}k_{(2)}m_{(2)}} \biggr) \\
			&\quad\quad\quad\quad+
			\sum_{i_{(1)}=1}^{N_{(1)}} v_{i_{(1)}}\sum_{j_{(1)}=1}^{N_{(1)}} \sum_{k_{(2)},m_{(2)},n_{(2)}=1}^{N_{(2)}} M_{k_{(2)}m_{(2)}n_{(2)}}^{i_{(1)}j_{(1)}} \biggr( \upsilon_{i_{(1)}k_{(2)}m_{(2)}n_{(2)}} -\upsilon_{j_{(1)}k_{(2)}m_{(2)}n_{(2)}} \biggr) \biggr) \\
			&\quad\quad\quad\quad+
			\mathcal{O}(\delta^2).
		\end{aligned}
	\end{equation}
	Specific details are presented in the Methods section.
	
	According to the condition $E_{\lambda}[\rho_{C_{(1)} \times C_{(2)}}+\rho_{C_{(1)} \times D_{(2)}}({\eta})] > E_{\lambda}^{\circ}[\rho_{C_{(1)} \times C_{(2)}}+\rho_{C_{(1)} \times D_{(2)}}({\eta})] $, we can explain whether population structure promotes cooperative behavior to replace betrayal behavior in $l_{(1)}$ network, the $l_{(2)}$ network is similar\citep{mcavoy2021fixation}. Based on equation $(11)$, we have calculated $(\frac{b}{c})^{*}$, critical benefit-to-cost ratio,for several complex two-layer networks, we obtain population structure promotes cooperative behavior to replace betrayal behavior under weak selection when condition $\frac{b}{c} > (\frac{b}{c})^{*} > 0$ is met\citep{mcavoy2021fixation}.
	
	We could calculate $(\frac{b}{c})^{*}$ of each layer of the network under condition ordinary initialization when the game between individuals is two-order. Similarly, we could calculate $(\frac{b}{c})^{*}$ of each layer of the network under condition ordinary initialization when the game between individuals is higher-order.We can conclude that in some networks, such as some ER stochastic networks, with two-order, when the bias coefficient approaches 0, the critical value $(\frac{b}{c})^{*}$ of the two-layer network is significantly smaller than that of the single-layer network. Therefore,the optimal bias for cooperative behavior in some tow-layer networks we obtain is $\alpha \rightarrow 0$, that is, one layer of network is completely dominated by another layer of network.Here are some examples display. 
	
	The four two-layer networks are the randomly generated ER two-layer network, which ${N_{(1)}}={N_{(2)}}=25$, the proportion of individuals associated with the other layer network is 1.0, the connection strength of these individuals is randomly generated,the bias coefficient is 0.1, 0.3, 0.6, 0.9 respectively, when the game between individuals is two-order.

	\begin{figure}[htbp]
		\centering
		\begin{minipage}[b]{0.45\textwidth}
			\centering
			\includegraphics[width=\textwidth]{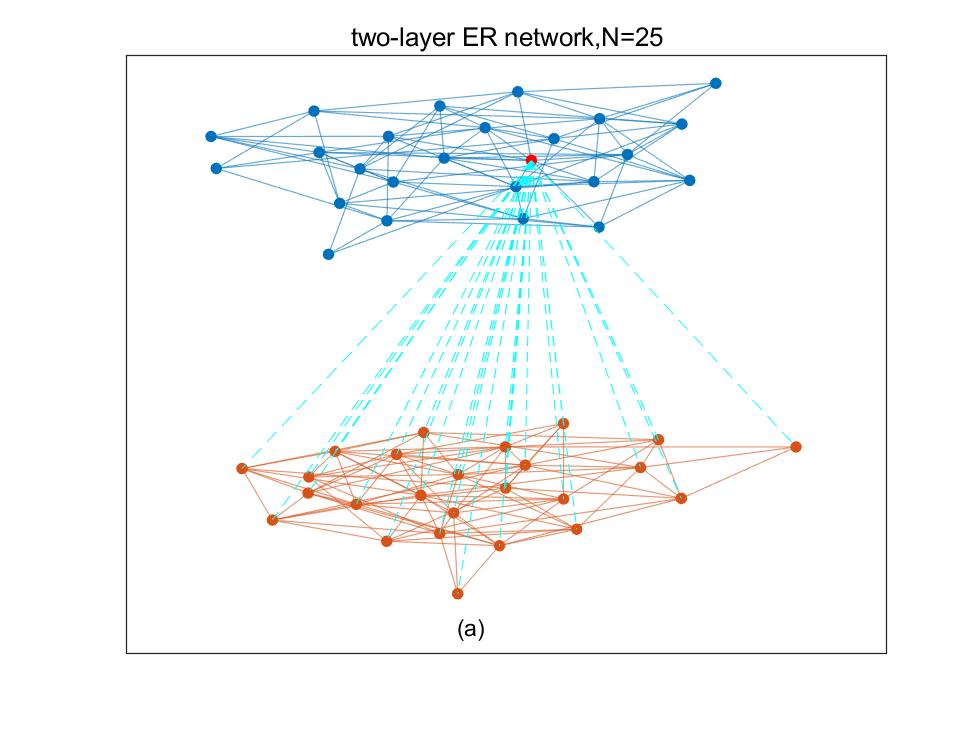}
		\end{minipage}
		\hfill
		\begin{minipage}[b]{0.45\textwidth}
			\centering
			\includegraphics[width=\textwidth]{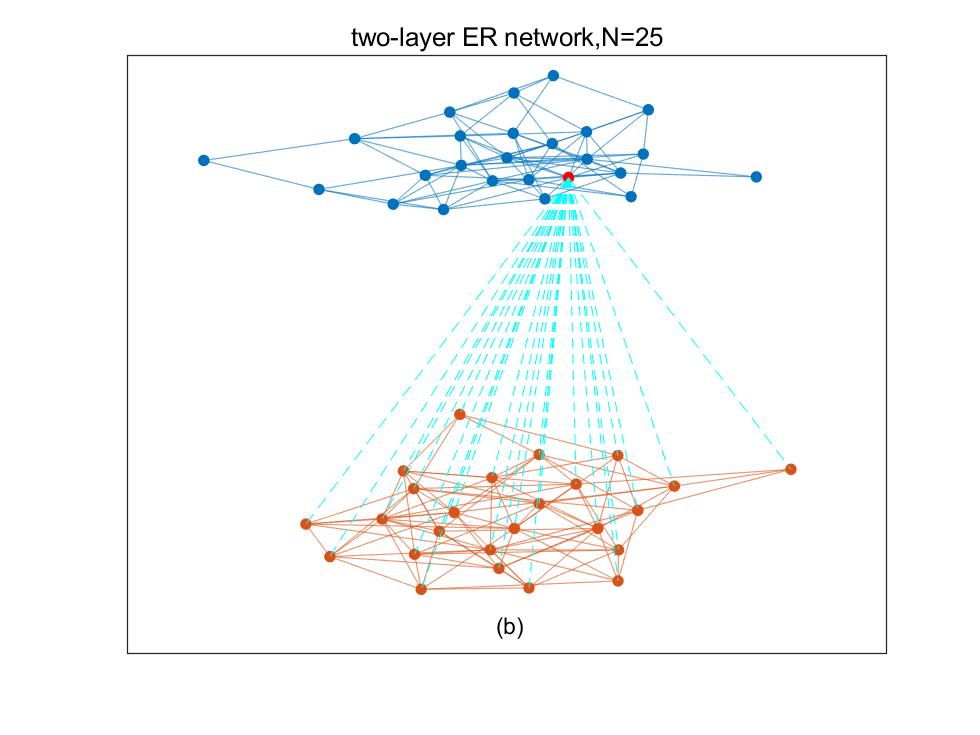}
		\end{minipage}
		
		\vspace{0.5cm} 
		
		\begin{minipage}[b]{0.45\textwidth}
			\centering
			\includegraphics[width=\textwidth]{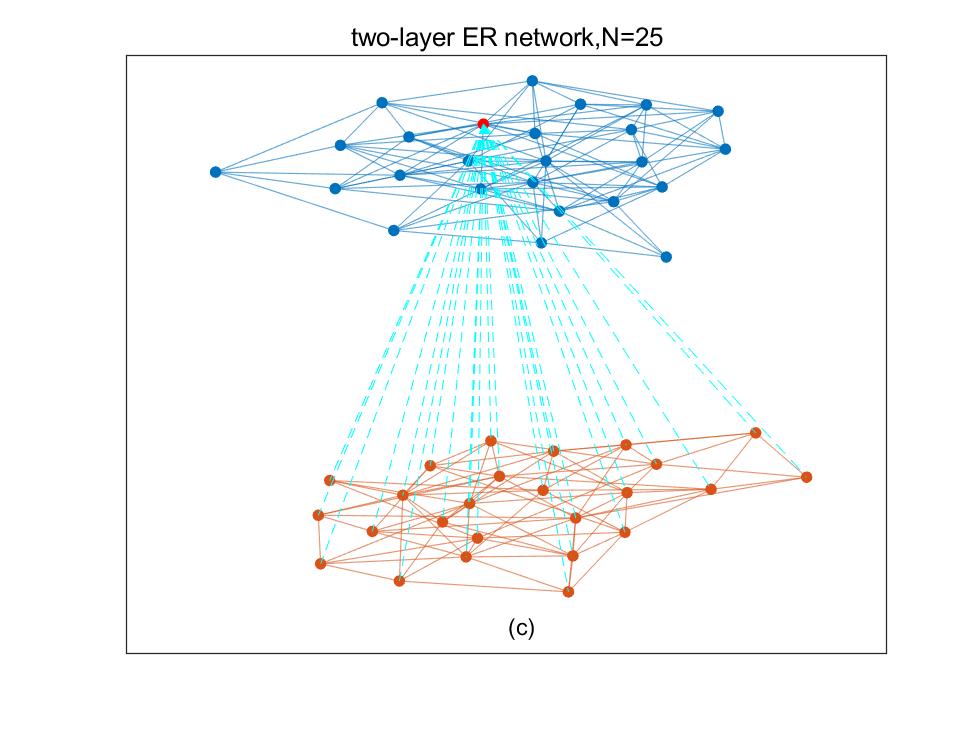}
		\end{minipage}
		\hfill
		\begin{minipage}[b]{0.45\textwidth}
			\centering
			\includegraphics[width=\textwidth]{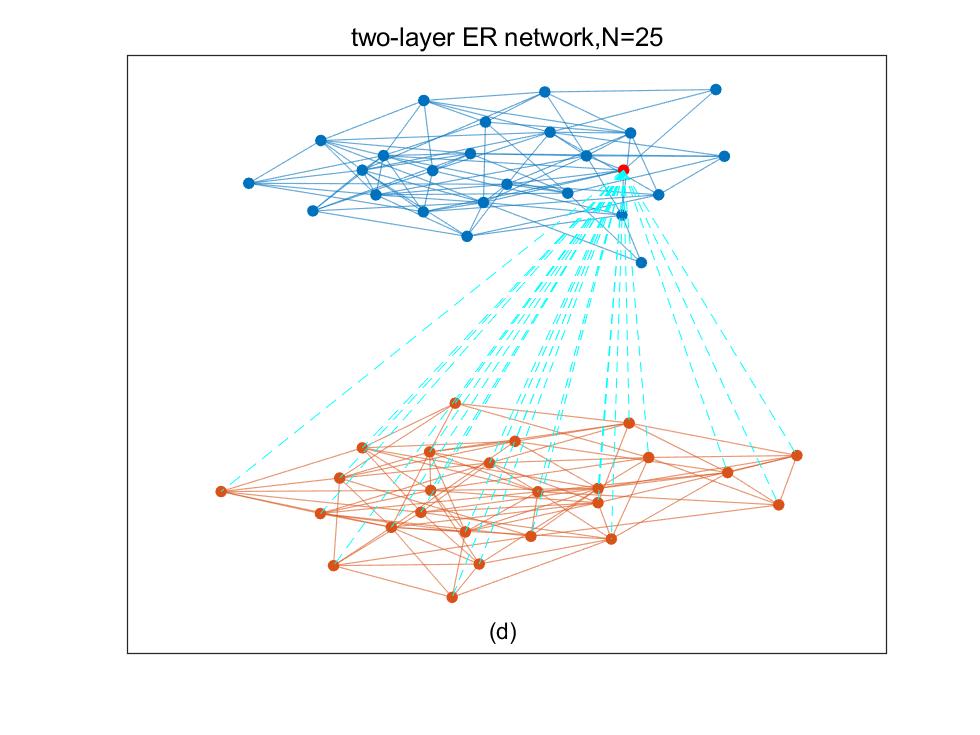}
		\end{minipage}
		
		\captionsetup{font=bf}
		\captionsetup{font=normalfont}
		\caption{\footnotesize the $(\frac{b}{c})^{*}$ of two-order game in two-layer ER network. (a): the critical value calculation result for the first type of network is $(\frac{b}{c})_{(1)}^{*}\approx26.67$, $(\frac{b}{c})_{(2)}^{*}\approx26.79$ when bias coefficient $\alpha=0.1$, $(\frac{b}{c})_{(1)}^{*}\approx29.37$, $(\frac{b}{c})_{(2)}^{*}\approx30.57$ when bias coefficient $\alpha=0.3$, $(\frac{b}{c})_{(1)}^{*}\approx31.39$, $(\frac{b}{c})_{(2)}^{*}\approx30.63$ when bias coefficient $\alpha=0.6$, $\frac{b}{c})_{(1)}^{*}\approx33.27$, $(\frac{b}{c})_{(2)}^{*}\approx33.69$ when bias coefficient $\alpha=0.9$. (b): the critical value calculation result for the second type of network is $(\frac{b}{c})_{(1)}^{*}\approx28.36$, $(\frac{b}{c})_{(2)}^{*}\approx28.99$ when bias coefficient $\alpha=0.1$, $(\frac{b}{c})_{(1)}^{*}\approx28.67$, $(\frac{b}{c})_{(2)}^{*}\approx29.56$ when bias coefficient $\alpha=0.3$, $(\frac{b}{c})_{(1)}^{*}\approx29.25$, $(\frac{b}{c})_{(2)}^{*}\approx29.57$ when bias coefficient $\alpha=0.6$, $\frac{b}{c})_{(1)}^{*}\approx30.29$, $(\frac{b}{c})_{(2)}^{*}\approx30.67$ when bias coefficient $\alpha=0.9$. (c): the critical value calculation result for the third type of network is $(\frac{b}{c})_{(1)}^{*}\approx25.27$, $(\frac{b}{c})_{(2)}^{*}\approx25.23$ when bias coefficient $\alpha=0.1$, $(\frac{b}{c})_{(1)}^{*}\approx27.66$, $(\frac{b}{c})_{(2)}^{*}\approx27.89$ when bias coefficient $\alpha=0.3$, $(\frac{b}{c})_{(1)}^{*}\approx28.29$, $(\frac{b}{c})_{(2)}^{*}\approx28.56$ when bias coefficient $\alpha=0.6$, $\frac{b}{c})_{(1)}^{*}\approx28.33$, $(\frac{b}{c})_{(2)}^{*}\approx28.97$ when bias coefficient $\alpha=0.9$. (d): the critical value calculation result for the fourth type of network is $(\frac{b}{c})_{(1)}^{*}\approx29.33$, $(\frac{b}{c})_{(2)}^{*}\approx29.36$ when bias coefficient $\alpha=0.1$, $(\frac{b}{c})_{(1)}^{*}\approx30.99$, $(\frac{b}{c})_{(2)}^{*}\approx30.69$ when bias coefficient $\alpha=0.3$, $(\frac{b}{c})_{(1)}^{*}\approx31.36$, $(\frac{b}{c})_{(2)}^{*}\approx31.25$ when bias coefficient $\alpha=0.6$, $\frac{b}{c})_{(1)}^{*}\approx31.56$, $(\frac{b}{c})_{(2)}^{*}\approx31.99$ when bias coefficient $\alpha=0.9$}.
	\end{figure}
	
	We can conclude that in two-order game and higher-order game, when the nonlinear factor satisfies certain conditions, some two-layer networks with profit feedback correlation is more conducive to cooperation than some single-layer networks when $\alpha \rightarrow 0$ under the same conditions.
	\section{Methods}
	We assume that the population completes corresponding policy updates at each time step according to specific update rules. We use $(Q_{(1)},f_{(1)})$ to represent the update event in the $l_{(1)}$ network\citep{su2022evolution}\citep{mcavoy2021fixation}, where $Q_{(1)}\subseteq{\{1_{(1)},...,N_{(1)}\}}$ represents a set of update individuals in the $l_{(1)}$ network, $f_{(1)}$: $Q_{(1)}\rightarrow{\{1_{(1)},...,N_{(1)}\}}$ represents the relationship between offspring and parent \citep{mcavoy2021fixation} in the $l_{(1)}$ network, similarly, a update event in the $l_{(2)}$ network, $(Q_{(2)},f_{(2)})$, where $Q_{(2)}\subseteq{\{1_{(2)},...,N_{(2)}\}}$ represents a set of update individuals in the $l_{(2)}$ network, $f_{(2)}$: $Q_{(2)}\rightarrow{\{1_{(2)},...,N_{(2)}\}}$ represents the relationship between offspring and parent in the $l_{(2)}$ network. According to the definition of update event in $l_{(1)}$ network and $l_{(2)}$ network, we obtain the expression of state of population at time $t+1$\citep{mcavoy2021fixation}\citep{su2022evolution}
	\begin{equation}\label{eq:example}
		s_{i(1)}^{t+1} =
		\begin{cases}
			s_{f_{(1)}(i_{(1)})}^{t} &  {i_{(1)}}\in{Q_{(1)}} \\
			s_{i(1)}^{t} &  {i_{(1)}}\notin{Q_{(1)}}.
		\end{cases}
	\end{equation}
	\begin{equation}\label{eq:example}
		s_{i(2)}^{t+1} =
		\begin{cases}
			s_{f_{(2)}(i_{(2)})}^{t} &  {i_{(2)}}\in{Q_{(2)}} \\
			s_{i(2)}^{t} &  {i_{(2)}}\notin{Q_{(2)}}.
		\end{cases}
	\end{equation}
	
	To facilitate the expression of this transformation, we introduce the promoted mapping in the $l_{(1)}$ network and the $l_{(2)}$ network\citep{mcavoy2021fixation}
	\begin{equation}\label{eq:example}
		\tilde{f}_{(1)}(i_{(1)}) =
		\begin{cases}
			f_{(1)}(i_{(1)}) &  {i_{(1)}}\in{Q_{(1)}} \\
			i_{(1)} &  {i_{(1)}}\notin{Q_{(1)}},
		\end{cases}
	\end{equation}
	\begin{equation}\label{eq:example}
		\tilde{f}_{(2)}(i_{(2)}) =
		\begin{cases}
			f_{(2)}(i_{(2)}) & {i_{(2)}}\in{Q_{(2)}}    \\
			i_{(2)}          & {i_{(2)}}\notin{Q_{(2)}}.
		\end{cases}
	\end{equation}
	
	We then obtain $\textbf{s}_{(1)}^{t+1}=\textbf{s}_{\tilde{f}_{(1)}}^{t}$,$\textbf{s}_{(2)}^{t+1}=\textbf{s}_{\tilde{f}_{(2)}}^{t}$,$\textbf{s}^{t+1} = \textbf{s}_{\tilde{f}_{(1)}}^{t} \times \textbf{s}_{\tilde{f}_{(2)}}^{t}$, we define $\textbf{s}_{\tilde{f}_{(1)} \times \tilde{f}_{(2)}}^{t}:=\textbf{s}_{\tilde{f}_{(1)}}^{t} \times \textbf{s}_{\tilde{f}_{(2)}}^{t}$\citep{su2022evolution}, we have $\textbf{s}^{t+1}=\textbf{s}_{\tilde{f}_{(1)} \times \tilde{f}_{(2)}}^{t}$. In state $\textbf{s}$, we denote the probability of update event occurring as $p_{(Q_{(1)},f_{(1)})}(\textbf{s})$,$p_{(Q_{(2)},f_{(2)})}(\textbf{s})$.We use  $P_{\textbf{s}\rightarrow\textbf{y}}$ to represent the probability of transferring from state $\textbf{s}$ to state $\textbf{y}$ in the Markov chain\citep{mcavoy2021fixation}.
	
	$\textbf{Example:}$ Moran process in tow-layer network\citep{mcavoy2021fixation}:
	\begin{equation}\label{eq:example}
		p_{(Q_{(1)},f_{(1)})}(\textbf{s}) =
		\begin{cases}
			{\frac{1}{N_{(1)}}}\frac{{F_{\tilde{f}_{(1)}({i_{(1)}})}}(\textbf{s}){w_{i_{(1)}{\tilde{f}_{(1)}({i_{(1)}})}}}}{\sum_{l_{(1)}=1}^{N_{(1)}} {F_{l_{(1)}}}(\textbf{s}){w_{l_{(1)}i_{(1)}}}} &  Q_{(1)}={i_{(1)}}\\
			0 &  |Q_{(1)}|\neq1.
		\end{cases}
	\end{equation}
	\begin{equation}\label{eq:example}
		p_{(Q_{(2)},f_{(2)})}(\textbf{s}) =
		\begin{cases}
			{\frac{1}{N_{(2)}}}\frac{{F_{\tilde{f}_{(2)}({i_{(2)}})}}(\textbf{s}){w_{i_{(2)}{\tilde{f}_{(2)}({i_{(2)}})}}}}{\sum_{l_{(2)}=1}^{N_{(2)}} {F_{l_{(2)}}}(\textbf{s}){w_{l_{(2)}i_{(2)}}}} &  Q_{(2)}={i_{(2)}}\\
			0 &  |Q_{(2)}|\neq1.
		\end{cases}
	\end{equation}
	
	We have provided the definition of update event. Since individuals policy update occurring only within the same layer, then we could obtain\citep{mcavoy2021fixation},
	\begin{equation}\label{eq:example}
		e_{i_{(1)}j_{(1)}}(\textbf{s}):=\sum_{\substack{(Q_{(1)},f_{(1)})\\j_{(1)}\in Q_{(1)},\tilde{f}_{(1)}(j_{(1)})=i_{(1)}}} p_{(Q_{(1)},f_{(1)})}(\textbf{s}),
	\end{equation}
	and
	\begin{equation}\label{eq:example}
		e_{i_{(2)}j_{(2)}}(\textbf{s}):=\sum_{\substack{(Q_{(2)},f_{(2)})\\j_{(2)}\in Q_{(2)},\tilde{f}_{(2)}(j_{(2)})=i_{(2)}}} p_{(Q_{(2)},f_{(2)})}(\textbf{s}).
	\end{equation}
	
	We introduce a pseudo-Boolean function to illustrate that  $\frac{d}{d\delta} \bigg|_{\delta=0}e_{i_{(1)}j_{(1)}}(\textbf{s})$ and $\frac{d}{d\delta} \bigg|_{\delta=0}e_{i_{(2)}j_{(2)}}(\textbf{s})$, we obtain $\frac{d}{d\delta} \bigg|_{\delta=0}e_{i_{(1)}j_{(1)}}(\textbf{s})$ and $\frac{d}{d\delta} \bigg|_{\delta=0}e_{i_{(2)}j_{(2)}}(\textbf{s})$ have unique polynomial functions in the two-layer network are\citep{mcavoy2021fixation}
	\begin{equation}\label{eq:example}
		\frac{d}{d\delta} \bigg|_{\delta=0}
		e_{i_{(1)}j_{(1)}}(\textbf{s}):=\sum_{\textbf{I}_{(1)}\subseteq {1_{(1)},...,N_{(1)}}} M_{\textbf{I}_{(1)}}^{i_{(1)}j_{(1)}}\textbf{s}_{\textbf{I}_{(1)}} + \sum_{\textbf{I}_{(2)}\subseteq {1_{(2)},...,N_{(2)}}} M_{\textbf{I}_{(2)}}^{i_{(1)}j_{(1)}}\textbf{s}_{\textbf{I}_{(2)}},
	\end{equation}
	and
	\begin{equation}\label{eq:example}
		\frac{d}{d\delta} \bigg|_{\delta=0}
		e_{i_{(2)}j_{(2)}}(\textbf{s}):=\sum_{\textbf{I}_{(2)}\subseteq {1_{(2)},...,N_{(2)}}} M_{\textbf{I}_{(2)}}^{i_{(2)}j_{(2)}}\textbf{s}_{\textbf{I}_{(2)}} + \sum_{\textbf{I}_{(1)}\subseteq {1_{(1)},...,N_{(1)}}} M_{\textbf{I}_{(1)}}^{i_{(2)}j_{(2)}}\textbf{s}_{\textbf{I}_{(1)}},
	\end{equation}
	where $M_{\textbf{I}_{(1)}}^{i_{(1)}j_{(1)}}$ is a real number that depends on $i_{(1)}$, $j_{(1)}$ and $\textbf{I}_{(1)}$, $M_{\textbf{I}_{(2)}}^{i_{(1)}j_{(1)}}$, $M_{\textbf{I}_{(2)}}^{i_{(2)}j_{(2)}}$and $M_{\textbf{I}_{(1)}}^{i_{(2)}j_{(2)}}$ are also similar.These real values not only reflect the key difference in the evolutionary dynamics of cooperation between the two-layer network and the single-layer network, but also quantify the impact of states $\textbf{s}$ on the probability of individual $j_{(1)}$ imitating individual $i_{(1)}$ strategy in $l_{(1)}$ network and the probability of individual $j_{(2)}$ imitating individual $i_{(2)}$ strategy in $l_{(2)}$ network.
	
	According to $e_{i_{(1)}j_{(1)}}(\textbf{s})$ and $e_{i_{(2)}j_{(2)}}(\textbf{s})$, the birth rate of $i_{(1)}$ is $b_{i_{(1)}}(\textbf{s})=\sum_{j_{(1)}=1}^{N_{(1)}} e_{i_{(1)}j_{(1)}}(\textbf{s})$ and the birth rate of $i_{(2)}$ is $b_{i_{(2)}}(\textbf{s})=\sum_{j_{(2)}=1}^{N_{(2)}} e_{i_{(2)}j_{(2)}}(\textbf{s})$\citep{mcavoy2021fixation}. The death rate of $i_{(1)}$ is $d_{i_{(1)}}(\textbf{s})=\sum_{j_{(1)}=1}^{N_{(1)}} e_{j_{(1)}i_{(1)}}(\textbf{s})$ and the birth rate of $i_{(2)}$ is $d_{i_{(2)}}(\textbf{s})=\sum_{j_{(2)}=1}^{N_{(2)}} e_{j_{(2)}i_{(2)}}(\textbf{s})$\citep{mcavoy2021fixation}.In the state $\textbf{s}$,we use the symbol $\Omega_{(1)}(\textbf{s})$, $\Omega_{(2)}(\textbf{s})$, to represent the expected change in the frequency of cooperative individuals in $l_{(1)}$ network and $l_{(2)}$ network\citep{mcavoy2021fixation}.
	\begin{equation}\label{eq:example}
		\Omega_{(1)}(\textbf{s}):=\sum_{i_{(1)}=1}^{N_{(1)}} s_{i_{(1)}}(b_{i_{(1)}}(\textbf{s})-d_{i_{(1)}}(\textbf{s})).
	\end{equation}
	\begin{equation}\label{eq:example}
		\Omega_{(2)}(\textbf{s}):=\sum_{i_{(2)}=1}^{N_{(2)}} s_{i_{(2)}}(b_{i_{(2)}}(\textbf{s})-d_{i_{(2)}}(\textbf{s})).
	\end{equation}
	
	We introduce $v_{i_{(1)}}$ and $v_{i_{(2)}}$ in the Result section\citep{mcavoy2021fixation}, then we update the definition of birth rates and death rates of $i_{(1)}$ to be $\hat{b}_{i_{(1)}}(\textbf{s}):=\sum_{j_{(1)}=1}^{N_{(1)}} e_{i_{(1)}j_{(1)}}(\textbf{s})v_{j_{(1)}}$ and $\hat{d}_{i_{(1)}}(\textbf{s}):=\sum_{j_{(1)}=1}^{N_{(1)}} e_{j_{(1)}i_{(1)}}(\textbf{s})v_{i_{(1)}}$,m $\hat{b}_{i_{(2)}}(\textbf{s}):=\sum_{j_{(2)}=1}^{N_{(2)}} e_{i_{(2)}j_{(2)}}(\textbf{s})v_{j_{(2)}}$ and $\hat{d}_{i_{(2)}}(\textbf{s}):=\sum_{j_{(2)}=1}^{N_{(2)}} e_{j_{(2)}i_{(2)}}(\textbf{s})v_{i_{(2)}}$. In the state $\textbf{s}$,we have\citep{mcavoy2021fixation}
	\begin{equation}\label{eq:example}
		\begin{aligned}
			&\hat{\Omega}_{(1)}(\textbf{s}) =
			&= \sum_{i_{(1)}=1}^{N_{(1)}} v_{i_{(1)}}\sum_{j_{(1)}=1}^{N_{(1)}} (s_{j_{(1)}}-s_{i_{(1)}})e_{j_{(1)}i_{(1)}},
		\end{aligned} 
	\end{equation}
	\begin{equation}\label{eq:example}
		\begin{aligned}
			&\hat{\Omega}_{(2)}(\textbf{s}) =
			&= \sum_{i_{(2)}=1}^{N_{(2)}} v_{i_{(2)}}\sum_{j_{(2)}=1}^{N_{(2)}} (s_{j_{(2)}}-s_{i_{(2)}})e_{j_{(2)}i_{(2)}}.
		\end{aligned} 
	\end{equation}
	we have $\hat{b}_{i_{(1)}}^{\circ}=\hat{d}_{i_{(1)}}^{\circ},\hat{b}_{i_{(2)}}^{\circ}=\hat{d}_{i_{(2)}}^{\circ}$, $\Omega_{sel}^{\circ}(\textbf{s})=0$.
	
	To calculate fixation probability, $\rho_{C_{(1)} \times C_{(2)}}(\textbf{$\eta$})$, it is useful to introduce the extended Markov chain and the sojourn times\citep{mcavoy2021fixation}\citep{mcavoy2020social}\citep{su2022evolution}. First, we introduce the extended Markov chain, we innovate a new parameter $\mu>0$, which could be understood as measuring the size of "mutation". Then we obtain extended averted probability of the extended Markov chain, for state $\textbf{s},\textbf{y} \in {{\mathbb{A}}^{N_{(1)}}} \times {{\mathbb{A}}^{N_{(2)}}}$,$\bm{\eta} \in {{\mathbb{A}}_{T}^{N_{(1)}} \times {\mathbb{A}}_{T}^{N_{(2)}}}$\citep{su2022evolution}\citep{mcavoy2021fixation}
	\begin{equation}\label{eq:example}
		P_{\textbf{s}\rightarrow\textbf{y}}^{ame(\bm{\eta})} =
		\begin{cases}
			\mu &  \textbf{s}\in\{{{\textbf{C}}_{(1)}} \times {{\textbf{C}}_{(2)}},{{\textbf{C}}_{(1)}} \times {{\textbf{D}}_{(2)}},{{\textbf{D}}_{(1)}} \times {{\textbf{C}}_{(2)}},{{\textbf{D}}_{(1)}} \times {{\textbf{D}}_{(2)}}\},\textbf{y}=\bm{\eta}\\
			1-\mu &  \textbf{s}=\textbf{y}={{\textbf{C}}_{(1)}} \times {{\textbf{C}}_{(2)}}\\
			1-\mu &  \textbf{s}=\textbf{y}={{\textbf{C}}_{(1)}} \times {{\textbf{D}}_{(2)}}\\
			1-\mu &  \textbf{s}=\textbf{y}={{\textbf{D}}_{(1)}} \times {{\textbf{C}}_{(2)}}\\
			1-\mu &  \textbf{s}=\textbf{y}={{\textbf{D}}_{(1)}} \times {{\textbf{D}}_{(2)}}\\
			P_{\textbf{s}\rightarrow\textbf{y}} &  \textbf{s}\notin\{{{\textbf{C}}_{(1)}} \times {{\textbf{C}}_{(2)}},{{\textbf{C}}_{(1)}} \times {{\textbf{D}}_{(2)}},{{\textbf{D}}_{(1)}} \times {{\textbf{C}}_{(2)}},{{\textbf{D}}_{(1)}} \times {{\textbf{D}}_{(2)}}\}.
		\end{cases}
	\end{equation}
	The extended Markov chain is provided with a unique stationary distribution\citep{fudenberg2006imitation}, we use $\{\pi_{ame(\bm{\eta})}(\textbf{s})\}_{\textbf{s}\in{{\mathbb{A}}^{N_{(1)}}} \times {\mathbb{A}}^{N_{(2)}}}$ to represent the unique stationary distribution. We could draw the following conclusion as\citep{su2022evolution}\citep{mcavoy2021fixation}
	\begin{equation}\label{eq:example}
		\lim\limits_{\mu \to 0}\pi_{ame(\bm{\eta})}(\textbf{s}) =
		\begin{cases}
			\rho_{C_{(1)} \times C_{(2)}}(\bm{\eta}) &  \textbf{s}={{\textbf{C}}_{(1)}} \times {{\textbf{C}}_{(2)}}\\
			\rho_{C_{(1)} \times D_{(2)}}(\bm{\eta}) &  \textbf{s}={{\textbf{C}}_{(1)}} \times {{\textbf{D}}_{(2)}}\\
			\rho_{D_{(1)} \times C_{(2)}}(\bm{\eta}) &  \textbf{s}={{\textbf{D}}_{(1)}} \times {{\textbf{C}}_{(2)}}\\
			\rho_{D_{(1)} \times D_{(2)}}(\bm{\eta}) &  \textbf{s}={{\textbf{D}}_{(1)}} \times {{\textbf{D}}_{(2)}}\\
			0 &  \textbf{s}\notin\{{{\textbf{C}}_{(1)}} \times {{\textbf{C}}_{(2)}},{{\textbf{C}}_{(1)}} \times {{\textbf{D}}_{(2)}},{{\textbf{D}}_{(1)}} \times {{\textbf{C}}_{(2)}},{{\textbf{D}}_{(1)}} \times {{\textbf{D}}_{(2)}}\}
		\end{cases}
	\end{equation}
	
	Secondly, we introduce the sojourn times\citep{mcavoy2021fixation}, $t_{\bm{\eta}}(\textbf{s})$, that indicate the expected number of get $\textbf{s}$ before reach to $\{{{\textbf{C}}_{(1)}} \times {{\textbf{C}}_{(2)}},{{\textbf{C}}_{(1)}} \times {{\textbf{D}}_{(2)}},{{\textbf{D}}_{(1)}} \times {{\textbf{C}}_{(2)}},{{\textbf{D}}_{(1)}} \times {{\textbf{D}}_{(2)}}\}$ when process's initial state is $\bm{\eta}$, for $\textbf{s}\in{{\mathbb{A}}^{N_{(1)}} \times {\mathbb{A}}^{N_{(2)}}}$,$\bm{\eta} \in {{\mathbb{A}}_{T}^{N_{(1)}} \times {\mathbb{A}}_{T}^{N_{(2)}}}$
	\begin{equation}\label{eq:example}
		t_{\bm{\eta}}(\textbf{s}) =
		\begin{cases}
			0 &  \textbf{s}\in\{{{\textbf{C}}_{(1)}} \times {{\textbf{C}}_{(2)}},{{\textbf{C}}_{(1)}} \times {{\textbf{D}}_{(2)}},{{\textbf{D}}_{(1)}} \times {{\textbf{C}}_{(2)}},{{\textbf{D}}_{(1)}} \times {{\textbf{D}}_{(2)}}\}\\
			1+\sum_{\textbf{y}\in{{\mathbb{A}}^{N_{(1)}} \times {\mathbb{A}}^{N_{(2)}}}} t_{\bm{\eta}}(\textbf{y})P_{\textbf{y}\rightarrow\textbf{s}}&  \textbf{s}=\bm{\eta}\\
			\sum_{\textbf{y}\in{{\mathbb{A}}^{N_{(1)}} \times {\mathbb{A}}^{N_{(2)}}}} t_{\bm{\eta}}(\textbf{y})P_{\textbf{y}\rightarrow\textbf{s}} &  \textbf{s}\notin\{{{\textbf{C}}_{(1)}} \times {{\textbf{C}}_{(2)}},{{\textbf{C}}_{(1)}} \times {{\textbf{D}}_{(2)}},{{\textbf{D}}_{(1)}} \times {{\textbf{C}}_{(2)}},{{\textbf{D}}_{(1)}} \times {{\textbf{D}}_{(2)}},\bm{\eta}\}.
		\end{cases}
	\end{equation}
	
	We let $\mathbb{B}$ denote the set ${\mathbb{A}}^{N_{(1)}} \times {\mathbb{A}}^{N_{(2)}}-\{{{\textbf{C}}_{(1)}} \times {{\textbf{C}}_{(2)}},{{\textbf{C}}_{(1)}} \times {{\textbf{D}}_{(2)}},{{\textbf{D}}_{(1)}} \times {{\textbf{C}}_{(2)}},{{\textbf{D}}_{(1)}} \times {{\textbf{D}}_{(2)}}\}$.
	
	Below we present some conclusions about fixation probability under weak selection in two-layer network\citep{su2022evolution}\citep{mcavoy2021fixation}.
	\begin{theorem}
		In the $l_{(1)}$ layer, the fixation probability of initial state $\bm{\eta}$ eventually transfer state ${\textbf{C}}_{(1)}$ under weak selection.
		\begin{equation}\label{eq:example}
			\rho_{{\textbf{C}}_{(1)} \times {\textbf{C}}_{(2)}}(\bm{\eta})+\rho_{{\textbf{C}}_{(1)} \times {\textbf{D}}_{(2)}}(\bm{\eta})=\hat{\eta}_{1}+\delta\Big\langle {\frac{d}{d\delta} \bigg|_{\delta=0}\hat{\Omega}_{1}} \Big\rangle_{\bm{\eta}}^{\circ}+\mathcal{O}(\delta^2).
		\end{equation}
	\end{theorem}
	$\textbf{Proof:}$ we could have the mathematical expression of expected change in reproductive—value-weighted frequency of $C$ in $l_{(1)}$ network under the amended Markov chain as follows\citep{su2022evolution}\citep{mcavoy2021fixation} 
	\begin{equation}\label{eq:example}
		\hat{\Omega}_{ame(\bm{\eta}){(1)}}(\textbf{s})=
		\begin{cases}
			\mu(\hat{\eta}_{(1)}-1) &  \textbf{s}={\textbf{C}}_{(1)} \times {\textbf{C}}_{(2)}\\
			\mu(\hat{\eta}_{(1)}-1) &  \textbf{s}={\textbf{C}}_{(1)} \times {\textbf{D}}_{(2)}\\
			\mu\hat{\eta}_{(1)} &  \textbf{s}={\textbf{D}}_{(1)} \times {\textbf{C}}_{(2)}\\
			\mu\hat{\eta}_{(1)} &  \textbf{s}={\textbf{D}}_{(1)} \times {\textbf{D}}_{(2)}\\
			\hat{\Omega}_{(1)}(\textbf{s})&  \textbf{s}\in\mathbb{B},
		\end{cases}
	\end{equation}
	we have\citep{su2022evolution}\citep{mcavoy2021fixation}
	\begin{equation}\label{eq:example}
		\begin{aligned}
			0 &= \mathbb{E}_{ame(\bm{\eta})}[\hat{\Omega}_{ame(\bm{\eta}){(1)}}]\\
			&= \mathbb{E}_{ame(\bm{\eta})}[\hat{\Omega}_{(1)}] \\
			&\quad+  \mu(\hat{\eta}_{1}-1)\pi_{ame(\bm{\eta})}({\textbf{C}}_{(1)} \times {\textbf{C}}_{(2)})+\mu(\hat{\eta}_{1}-1)\pi_{ame(\bm{\eta})}({\textbf{C}}_{(1)} \times {\textbf{D}}_{(2)}) \\
			&\quad+  \mu\hat{\eta}_{1}\pi_{ame(\bm{\eta})}({\textbf{D}}_{(1)} \times {\textbf{C}}_{(2)})+\mu\hat{\eta}_{1}\pi_{ame(\bm{\eta})}({\textbf{D}}_{(1)} \times {\textbf{D}}_{(2)}).
		\end{aligned}
	\end{equation}
	Take the derivative of both sides of the equation and follow equation(38),we have\citep{su2022evolution}\citep{mcavoy2021fixation}
	\begin{equation}\label{eq:example}
		\rho_{{\textbf{C}}_{(1)} \times {\textbf{C}}_{(2)}}(\bm{\eta})+\rho_{{\textbf{C}}_{(1)} \times {\textbf{D}}_{(2)}}(\bm{\eta})=\hat{\eta}_{1}+\Big\langle \hat{\Omega}_{1} \Big\rangle_{\bm{\eta}}
	\end{equation}
	then according to the weak selection,we can perform Taylor expansion at $\delta=0$,the expansion formula is as follows\citep{su2022evolution}\citep{mcavoy2021fixation}
	\begin{equation}\label{eq:example}
		\begin{aligned}
			\Big\langle \hat{\Omega}_{1} \Big\rangle_{\bm{\eta}} &= \delta{\frac{d}{d\delta} \bigg|_{\delta=0}\Big\langle \hat{\Omega}_{1}} \Big\rangle_{\bm{\eta}}^{\circ}+\mathcal{O}(\delta^2) \\
			&= \delta\Big\langle {\frac{d}{d\delta} \bigg|_{\delta=0}\hat{\Omega}_{1}} \Big\rangle_{\bm{\eta}}^{\circ}+\mathcal{O}(\delta^2).
		\end{aligned}
	\end{equation}
	
	Similarly,we have the theorem of the fixation probability of state $\bm{\eta}$ eventually transfer state ${\textbf{C}}_{(2)}$ under weak selection in the $l_{(2)}$ layer.
	Further simplify the formula, we can obtain the following corollary about the fixation probability of two-order game in the $l_{(1)}$ layer\citep{su2022evolution}\citep{mcavoy2021fixation}.
	\begin{corollary}
		\begin{equation}\label{eq:example}
			\begin{aligned}
				&\rho_{{\textbf{C}}_{(1)} \times {\textbf{C}}_{(2)}}(\bm{\eta})+\rho_{{\textbf{C}}_{(1)} \times {\textbf{D}}_{(2)}}(\bm{\eta})=\hat{\eta}_{(1)} \\
				&\quad\quad\quad\quad+
				\frac{\delta}{N_{(1)}}\biggl(\sum_{i_{(1)}=1}^{N_{(1)}} v_{i_{(1)}}\sum_{j_{(1)}=1}^{N_{(1)}} \sum_{k_{(1)}=1}^{N_{(1)}} M_{k_{(1)}}^{i_{(1)}j_{(1)}} \left( \psi_{i_{(1)}k_{(1)}} -\psi_{j_{(1)}k_{(1)}} \right)  \\
				&\quad\quad\quad\quad+
				\sum_{i_{(1)}=1}^{N_{(1)}} v_{i_{(1)}}\sum_{j_{(1)}=1}^{N_{(1)}} \sum_{k_{(2)}=1}^{N_{(2)}} M_{k_{(2)}}^{i_{(1)}j_{(1)}}\left( \phi_{i_{(1)}k_{(2)}}-\phi_{j_{(1)}k_{(2)}} \right) \biggr) \\
				&\quad\quad\quad\quad+
				\mathcal{O}(\delta^2),
			\end{aligned}
		\end{equation}
		where
		\begin{equation}\label{eq:example}
			\left\{
			\begin{aligned}
				&M_{k_{(1)}}^{i_{(1)}j_{(1)}} = \alpha{p_{i_{(1)}j_{(1)}}^{(1)}}\biggl(-c\biggl(p_{j_{(1)}k_{(1)}}^{(0)}w_{j_{(1)}} - p_{i_{(1)}k_{(1)}}^{(1)}w_{k_{(1)}}\biggr) \\
				&\quad\quad\quad\quad+  b\biggl(w_{j_{(1)}k_{(1)}} - \sum_{l_{(1)}=1}^{N_{(1)}} p_{i_{(1)}l_{(1)}}^{(1)}w_{l_{(1)}k_{(1)}}\biggr) \biggr) \\
				&M_{k_{(2)}}^{i_{(1)}j_{(1)}} = (1 - \alpha){p_{i_{(1)}j_{(1)}}^{(1)}}\biggl(-c\biggl(q_{j_{(1)}k_{(2)}}^{(0)}w_{k_{(2)}} - \sum_{l_{(1)}=1}^{N_{(1)}} p_{i_{(1)}l_{(1)}}^{(1)}q_{l_{(1)}k_{(2)}}w_{k_{(2)}}\biggr) \\
				&\quad\quad\quad\quad+  b\biggl(\sum_{n_{(2)}=1}^{N_{(2)}} q_{j_{(1)}n_{(2)}}w_{n_{(2)}k_{(2)}} - \sum_{l_{(1)}=1}^{N_{(1)}} p_{i_{(1)}l_{(1)}}^{(1)}\sum_{n_{(2)}=1}^{N_{(2)}} q_{l_{(1)}n_{(2)}}w_{n_{(2)}k_{(2)}}\biggr) \biggr), \\
			\end{aligned}
			\right.
		\end{equation}
		where
		\begin{equation}\label{eq:example}
			\psi_{i_{(1)}j_{(1)}} = \frac{N_{(1)}}{2}(\hat{\eta}_{(1)} - \eta_{i_{(1)}j_{(1)}}) + \frac{1}{2}\sum_{k_{(1)}=1}^{N_{(1)}} \left(p_{i_{(1)}k_{(1)}}^{(1)}\psi_{k_{(1)}j_{(1)}} + p_{j_{(1)}k_{(1)}}^{(1)}\psi_{i_{(1)}k_{(1)}}\right),
		\end{equation}
		\begin{equation}\label{eq:example}
			\begin{aligned}
				&\phi_{i_{(1)}j_{(2)}} = \frac{N_{(1)}N_{(2)}}{N_{(1)}+N_{(2)}-1}((\hat{\eta}_{(1)}\hat{\eta}_{(2)}) - \eta_{i_{(1)}j_{(2)}}) \\
				&\quad\quad\quad\quad+
				\frac{1}{N_{(1)}+N_{(2)}-1}\left(\sum_{k_{(1)}=1}^{N_{(1)}}\sum_{k_{(2)}=1}^{N_{(2)}} p_{i_{(1)}k_{(1)}}^{(1)} p_{j_{(2)}k_{(2)}}^{(1)}\phi_{k_{(1)}k_{(2)}}\right) \\
				&\quad\quad\quad\quad+
				\frac{N_{(2)}-1}{N_{(1)}+N_{(2)}-1}\sum_{k_{(1)}=1}^{N_{(1)}}p_{i_{(1)}k_{(1)}}^{(1)}\phi_{k_{(1)}j_{(2)}}+\frac{N_{(1)}-1}{N_{(1)}+N_{(2)}-1}\sum_{k_{(2)}=1}^{N_{(2)}}p_{j_{(2)}k_{(2)}}^{(1)}\phi_{i_{(1)}k_{(2)}},
			\end{aligned}
		\end{equation}
	\end{corollary}
	$\textbf{Proof:}$ According to the equation(1), we could have the corollary.
	
	Similarly, we have the corollary of the fixation probability of two-order game in the $l_{(2)}$ layer under weak selection\citep{su2022evolution}\citep{mcavoy2021fixation}.
	\begin{corollary}
		\begin{equation}\label{eq:example}
			\begin{aligned}
				&\rho_{{\textbf{C}}_{(1)} \times {\textbf{C}}_{(2)}}(\bm{\eta})+\rho_{{\textbf{D}}_{(1)} \times {\textbf{C}}_{(2)}}(\bm{\eta})=\hat{\eta}_{(2)} \\
				&\quad\quad\quad\quad+
				\frac{\delta}{N_{(2)}}\biggl(\sum_{i_{(2)}=1}^{N_{(2)}} v_{i_{(2)}}\sum_{j_{(2)}=1}^{N_{(2)}} \sum_{k_{(2)}=1}^{N_{(2)}} M_{k_{(2)}}^{i_{(2)}j_{(2)}}\left( \xi_{i_{(2)}k_{(2)}} -\xi_{j_{(2)}k_{(2)}}\right) \\
				&\quad\quad\quad\quad+
				\sum_{i_{(2)}=1}^{N_{(2)}} v_{i_{(2)}}\sum_{j_{(2)}=1}^{N_{(2)}} \sum_{k_{(1)}=1}^{N_{(1)}} M_{k_{(1)}}^{i_{(2)}j_{(2)}}\left( \zeta_{i_{(2)}k_{(1)}} -\zeta_{j_{(2)}k_{(1)}} \right)  \biggr)\\
				&\quad\quad\quad\quad+
				\mathcal{O}(\delta^2),
			\end{aligned}
		\end{equation}
		where
		\begin{equation}\label{eq:example}
			\left\{
			\begin{aligned}
				&M_{k_{(2)}}^{i_{(2)}j_{(2)}} = \alpha{p_{i_{(2)}j_{(2)}}^{(1)}}\biggl(-c\biggl(p_{j_{(2)}k_{(2)}}^{(0)}w_{j_{(2)}} - p_{i_{(2)}k_{(2)}}^{(1)}w_{k_{(2)}}\biggr) \\
				&\quad\quad\quad\quad+  b\biggl(w_{j_{(2)}k_{(2)}} - \sum_{l_{(2)}=1}^{N_{(2)}} p_{i_{(2)}l_{(2)}}^{(1)}w_{l_{(2)}k_{(2)}}\biggr) \biggr) \\
				&M_{k_{(1)}}^{i_{(2)}j_{(2)}} = (1 - \alpha){p_{i_{(2)}j_{(2)}}^{(1)}}\biggl(-c\biggl(q_{j_{(2)}k_{(1)}}^{(0)}w_{k_{(1)}} - \sum_{l_{(2)}=1}^{N_{(2)}} p_{i_{(2)}l_{(2)}}^{(1)}q_{l_{(2)}k_{(1)}}w_{k_{(1)}}\biggr) \\
				&\quad\quad\quad\quad+  b\biggl(\sum_{n_{(1)}=1}^{N_{(1)}} q_{j_{(2)}n_{(1)}}w_{n_{(1)}k_{(1)}} - \sum_{l_{(2)}=1}^{N_{(2)}} p_{i_{(2)}l_{(2)}}^{(1)}\sum_{n_{(1)}=1}^{N_{(1)}} q_{l_{(2)}n_{(1)}}w_{n_{(1)}k_{(1)}}\biggr) \biggr), \\
			\end{aligned}
			\right.
		\end{equation}
		where 
		\begin{equation}\label{eq:example}
			\xi_{i_{(2)}j_{(2)}} = \frac{N_{(2)}}{2}(\hat{\eta}_{(2)} - \eta_{i_{(2)}j_{(2)}}) + \frac{1}{2}\sum_{k_{(2)}=1}^{N_{(2)}} \left(p_{i_{(2)}k_{(2)}}^{(1)}\xi_{k_{(2)}j_{(2)}} + p_{j_{(2)}k_{(2)}}^{(1)}\xi_{i_{(2)}k_{(2)}}\right),
		\end{equation}
		\begin{equation}\label{eq:example}
			\begin{aligned}
				&\zeta_{i_{(2)}j_{(1)}} = \frac{N_{(1)}N_{(2)}}{N_{(1)}+N_{(2)}-1}((\hat{\eta}_{(1)}\hat{\eta}_{(2)}) - \eta_{i_{(2)}j_{(1)}}) \\
				&\quad\quad\quad\quad+
				\frac{1}{N_{(1)}+N_{(2)}-1}\left(\sum_{k_{(1)}=1}^{N_{(1)}}\sum_{k_{(2)}=1}^{N_{(2)}} p_{j_{(1)}k_{(1)}}^{(1)} p_{i_{(2)}k_{(2)}}^{(1)} \zeta_{k_{(2)}k_{(1)}}\right) \\
				&\quad\quad\quad\quad+
				\frac{N_{(2)}-1}{N_{(1)}+N_{(2)}-1}\sum_{k_{(1)}=1}^{N_{(1)}}p_{j_{(1)}k_{(1)}}^{(1)}\zeta_{k_{(1)}i_{(2)}}+\frac{N_{(1)}-1}{N_{(1)}+N_{(2)}-1}\sum_{k_{(2)}=1}^{N_{(2)}}p_{i_{(2)}k_{(2)}}^{(1)}\zeta_{j_{(1)}k_{(2)}}, \\
			\end{aligned}
		\end{equation}
	\end{corollary}
	
	Similarly,we have the corollary of the fixation probability of higher-order game in the $l_{(1)}$ layer under weak selection\citep{su2022evolution}\citep{mcavoy2021fixation}.
	\begin{corollary}
		\begin{equation}\label{eq:example}
			\begin{aligned}
				&\rho_{{\textbf{C}}_{(1)} \times {\textbf{C}}_{(2)}}(\bm{\eta})+\rho_{{\textbf{C}}_{(1)} \times {\textbf{D}}_{(2)}}(\bm{\eta}) =
				\hat{\eta}_{(1)}+ \frac{\delta}{N_{(1)}}\biggr(\sum_{i_{(1)}=1}^{N_{(1)}} v_{i_{(1)}}\sum_{j_{(1)}=1}^{N_{(1)}} \sum_{k_{(1)}=1}^{N_{(1)}} M_{k_{(1)}}^{i_{(1)}j_{(1)}} \biggr( \psi_{i_{(1)}k_{(1)}} -\psi_{j_{(1)}k_{(1)}} \biggr)  \\
				&\quad\quad\quad\quad+
				\sum_{i_{(1)}=1}^{N_{(1)}} v_{i_{(1)}}\sum_{j_{(1)}=1}^{N_{(1)}} \sum_{k_{(1)},m_{(1)}=1}^{N_{(1)}} M_{k_{(1)}m_{(1)}}^{i_{(1)}j_{(1)}} \biggr( \nu_{i_{(1)}k_{(1)}m_{(1)}} -\nu_{j_{(1)}k_{(1)}m_{(1)}} \biggr) \\
				&\quad\quad\quad\quad+
				\sum_{i_{(1)}=1}^{N_{(1)}} v_{i_{(1)}}\sum_{j_{(1)}=1}^{N_{(1)}} \sum_{k_{(1)},m_{(1)},n_{(1)}=1}^{N_{(1)}} M_{k_{(1)}m_{(1)}n_{(1)}}^{i_{(1)}j_{(1)}} \biggr( \kappa_{i_{(1)}k_{(1)}m_{(1)}n_{(1)}} -\kappa_{j_{(1)}k_{(1)}m_{(1)}n_{(1)}}\biggr) \\
				&\quad\quad\quad\quad+
				\sum_{i_{(1)}=1}^{N_{(1)}} v_{i_{(1)}}\sum_{j_{(1)}=1}^{N_{(1)}} \sum_{k_{(2)}=1}^{N_{(2)}} M_{k_{(2)}}^{i_{(1)}j_{(1)}}\biggr( \phi_{i_{(1)}k_{(2)}} -\phi_{j_{(1)}k_{(2)}} \biggr) \\
				&\quad\quad\quad\quad+
				\sum_{i_{(1)}=1}^{N_{(1)}} v_{i_{(1)}}\sum_{j_{(1)}=1}^{N_{(1)}} \sum_{k_{(2)},m_{(2)}=1}^{N_{(2)}} M_{k_{(2)}m_{(2)}}^{i_{(1)}j_{(1)}}\biggr( \sigma_{i_{(1)}k_{(2)}m_{(2)}} -\sigma_{j_{(1)}k_{(2)}m_{(2)}} \biggr) \\
				&\quad\quad\quad\quad+
				\sum_{i_{(1)}=1}^{N_{(1)}} v_{i_{(1)}}\sum_{j_{(1)}=1}^{N_{(1)}} \sum_{k_{(2)},m_{(2)},n_{(2)}=1}^{N_{(2)}} M_{k_{(2)}m_{(2)}n_{(2)}}^{i_{(1)}j_{(1)}} \biggr( \upsilon_{i_{(1)}k_{(2)}m_{(2)}n_{(2)}} -\upsilon_{j_{(1)}k_{(2)}m_{(2)}n_{(2)}} \biggr) \biggr) \\
				&\quad\quad\quad\quad+
				\mathcal{O}(\delta^2),
			\end{aligned}
		\end{equation}
		where
		\begin{equation}\label{eq:example}
			\begin{aligned}
				&M_{k_{(1)}}^{i_{(1)}j_{(1)}} = \alpha{p_{i_{(1)}j_{(1)}}^{(1)}}\biggl(-c\biggl(p_{j_{(1)}k_{(1)}}^{(0)}\sum_{n_{(1)}=1}^{N_{(1)}}w_{j_{(1)}n_{(1)}}+\frac{1}{2}p_{j_{(1)}k_{(1)}}^{(0)}\sum_{n_{(1)},m_{(1)}=1}^{N_{(1)}}w_{j_{(1)}n_{(1)}m_{(1)}} \\
				&\quad\quad\quad\quad-
				p_{i_{(1)}k_{(1)}}^{(1)}\sum_{n_{(1)}=1}^{N_{(1)}}w_{k_{(1)}n_{(1)}}-\frac{1}{2}p_{i_{(1)}k_{(1)}}^{(1)}\sum_{n_{(1)},m_{(1)}=1}^{N_{(1)}}w_{k_{(1)}n_{(1)}m_{(1)}}\biggr) \\
				&\quad\quad\quad\quad+  b\biggl(\frac{1}{2}p_{j_{(1)}k_{(1)}}^{(0)}\sum_{n_{(1)}=1}^{N_{(1)}}w_{j_{(1)}n_{(1)}}+\frac{1}{6}p_{j_{(1)}k_{(1)}}^{(0)}\sum_{n_{(1)},m_{(1)}=1}^{N_{(1)}}w_{j_{(1)}n_{(1)}m_{(1)}}-\frac{1}{2}p_{i_{(1)}k_{(1)}}^{(1)}\sum_{n_{(1)}=1}^{N_{(1)}}w_{k_{(1)}n_{(1)}} \\
				&\quad\quad\quad\quad+
				\frac{1}{2}w_{j_{(1)}k_{(1)}}+\frac{1}{6}\sum_{n_{(1)}=1}^{N_{(1)}}w_{j_{(1)}k_{(1)}n_{(1)}}+\frac{1}{6}\sum_{n_{(1)}=1}^{N_{(1)}}w_{j_{(1)}n_{(1)}k_{(1)}}
				&\quad\quad\quad\quad-
				\frac{1}{6}p_{i_{(1)}k_{(1)}}^{(1)}\sum_{n_{(1)},m_{(1)}=1}^{N_{(1)}}w_{k_{(1)}n_{(1)}m_{(1)}}-\frac{1}{2}\sum_{n_{(1)}=1}^{N_{(1)}}p_{i_{(1)}n_{(1)}}^{(1)}w_{n_{(1)}k_{(1)}}-\frac{1}{6}\sum_{n_{(1)}=1}^{N_{(1)}}p_{i_{(1)}n_{(1)}}^{(1)}\sum_{m_{(1)}=1}^{N_{(1)}}w_{n_{(1)}k_{(1)}m_{(1)}} \\
				&\quad\quad\quad\quad-
				\frac{1}{6}\sum_{n_{(1)}=1}^{N_{(1)}}p_{i_{(1)}n_{(1)}}^{(1)}\sum_{m_{(1)}=1}^{N_{(1)}}w_{n_{(1)}m_{(1)}k_{(1)}}-\frac{1}{6}\sum_{n_{(1)}=1}^{N_{(1)}}p_{i_{(1)}n_{(1)}}^{(1)}\sum_{m_{(1)}=1}^{N_{(1)}}w_{n_{(1)}k_{(1)}m_{(1)}} \\
				&\quad\quad\quad\quad-
				\frac{1}{2}\sum_{n_{(1)}=1}^{N_{(1)}}p_{i_{(1)}n_{(1)}}^{(1)}w_{n_{(1)}k_{(1)}}-\frac{1}{6}p_{i_{(1)}k_{(1)}}^{(1)}\sum_{n_{(1)},m_{(1)}=1}^{N_{(1)}}w_{k_{(1)}n_{(1)}m_{(1)}}\biggr) \biggr), \\
			\end{aligned}
		\end{equation}
		\begin{equation}\label{eq:example2}
			\begin{aligned}
				&M_{k_{(1)}m_{(1)}}^{i_{(1)}j_{(1)}} =
				\alpha{p_{i_{(1)}j_{(1)}}^{(1)}}\biggl(b\biggl(p_{j_{(1)}k_{(1)}}^{(0)}\biggl(\frac{\theta_{2}-1}{2}w_{j_{(1)}m_{(1)}}+\frac{\theta_{3}-1}{6}\sum_{h_{(1)}=1}^{N_{(1)}}w_{j_{(1)}m_{(1)}h_{(1)}} \\
				&\quad\quad\quad\quad+
				\frac{\theta_{3}-1}{6}\sum_{h_{(1)}=1}^{N_{(1)}}w_{j_{(1)}h_{(1)}m_{(1)}}\biggr)+\frac{\theta_{3}-1}{6}w_{j_{(1)}k_{(1)}m_{(1)}}-\frac{\theta_{2}-1}{2}p_{i_{(1)}k_{(1)}}^{(1)}w_{k_{(1)}m_{(1)}} \\
				&\quad\quad\quad\quad-
				\frac{\theta_{3}-1}{6}p_{i_{(1)}k_{(1)}}^{(1)}\sum_{n_{(1)}=1}^{N_{(1)}}w_{k_{(1)}m_{(1)}n_{(1)}}-\frac{\theta_{3}-1}{6}p_{i_{(1)}k_{(1)}}^{(1)}\sum_{n_{(1)}=1}^{N_{(1)}}w_{k_{(1)}n_{(1)}m_{(1)}} \\
				&\quad\quad\quad\quad-
				\frac{\theta_{3}-1}{6}\sum_{n_{(1)}=1}^{N_{(1)}}p_{i_{(1)}n_{(1)}}^{(1)}w_{n_{(1)}k_{(1)}m_{(1)}}\biggr) \biggr), \\
			\end{aligned}
		\end{equation}
		\begin{equation}\label{eq:example2}
			\begin{aligned}
				&M_{k_{(1)}m_{(1)}n_{(1)}}^{i_{(1)}j_{(1)}} =
				\alpha{p_{i_{(1)}j_{(1)}}^{(1)}}\biggl(b\biggl(p_{j_{(1)}k_{(1)}}^{(0)}\frac{{(\theta_{3}-1)}^{2}}{6}w_{j_{(1)}m_{(1)}n_{(1)}}-\frac{{(\theta_{3}-1)}^{2}}{6}p_{i_{(1)}k_{(1)}}^{(1)}w_{k_{(1)}m_{(1)}n_{(1)}}\biggr) \biggr), \\
			\end{aligned}
		\end{equation}
		\begin{equation}\label{eq:example2}
			\begin{aligned}
				&M_{k_{(2)}}^{i_{(1)}j_{(1)}} = (1-\alpha){p_{i_{(1)}j_{(1)}}^{(1)}}\biggl(-c\biggl(q_{j_{(1)}k_{(2)}}\sum_{h_{(2)}=1}^{N_{(2)}}w_{k_{(2)}h_{(2)}}+\frac{1}{2}q_{j_{(1)}k_{(2)}}\sum_{l_{(2)},r_{(2)}=1}^{N_{(2)}}w_{k_{(2)}l_{(2)}r_{(2)}} \\
				&\quad\quad\quad\quad-
				\sum_{u_{(1)}=1}^{N_{(1)}}p_{i_{(1)}u_{(1)}}^{(1)}\biggl(q_{u_{(1)}k_{(2)}}\sum_{h_{(2)}=1}^{N_{(2)}}w_{k_{(2)}h_{(2)}}+\frac{1}{2}q_{u_{(1)}k_{(2)}}\sum_{l_{(2)},r_{(2)}=1}^{N_{(2)}}w_{k_{(2)}l_{(2)}r_{(2)}}\biggr) \biggr) \\
				&\quad\quad\quad\quad+  b\biggr(\frac{1}{2}q_{j_{(1)}k_{(2)}}\sum_{h_{(2)}=1}^{N_{(2)}}w_{k_{(2)}h_{(2)}}+\frac{1}{2}\sum_{h_{(2)}=1}^{N_{(2)}}q_{j_{(1)}h_{(2)}}w_{h_{(2)}k_{(2)}}+\frac{1}{6}q_{j_{(1)}k_{(2)}}\sum_{l_{(2)},r_{(2)}=1}^{N_{(2)}}w_{k_{(2)}l_{(2)}r_{(2)}} \\
				&\quad\quad\quad\quad+ 
				\frac{1}{6}\sum_{h_{(2)}=1}^{N_{(2)}}q_{j_{(1)}h_{(2)}}\sum_{r_{(2)}=1}^{N_{(2)}}w_{h_{(2)}k_{(2)}r_{(2)}}+\frac{1}{6}\sum_{h_{(2)}=1}^{N_{(2)}}q_{j_{(1)}h_{(2)}}\sum_{r_{(2)}=1}^{N_{(2)}}w_{h_{(2)}r_{(2)}k_{(2)}} \\
				&\quad\quad\quad\quad-
				\sum_{u_{(1)}=1}^{N_{(1)}}p_{i_{(1)}u_{(1)}}^{(1)}\biggl(\frac{1}{2}q_{u_{(1)}k_{(2)}}\sum_{h_{(2)}=1}^{N_{(2)}}w_{k_{(2)}h_{(2)}}+\frac{1}{2}\sum_{h_{(2)}=1}^{N_{(2)}}q_{u_{(1)}h_{(2)}}w_{h_{(2)}k_{(2)}} \\
				&\quad\quad\quad\quad+ 
				\frac{1}{6}q_{u_{(1)}k_{(2)}}\sum_{l_{(2)},r_{(2)}=1}^{N_{(2)}}w_{k_{(2)}l_{(2)}r_{(2)}}+\frac{1}{6}\sum_{h_{(2)}=1}^{N_{(2)}}q_{u_{(1)}h_{(2)}}\sum_{r_{(2)}=1}^{N_{(2)}}w_{h_{(2)}k_{(2)}r_{(2)}} \\
				&\quad\quad\quad\quad+ 
				\frac{1}{6}\sum_{h_{(2)}=1}^{N_{(2)}}q_{u_{(1)}h_{(2)}}\sum_{r_{(2)}=1}^{N_{(2)}}w_{h_{(2)}r_{(2)}k_{(2)}}\biggr) \biggr) \biggr),\\
			\end{aligned}
		\end{equation}
		\begin{equation}\label{eq:example2}
			\begin{aligned}
				&M_{k_{(2)}m_{(2)}}^{i_{(1)}j_{(1)}} =
				(1-\alpha){p_{i_{(1)}j_{(1)}}^{(1)}}\biggl(b\biggl(\frac{\theta_{2}-1}{2}q_{j_{(1)}k_{(2)}}w_{k_{(2)}m_{(2)}}+\frac{\theta_{3}-1}{6}q_{j_{(1)}k_{(2)}}\sum_{r_{(2)}=1}^{N_{(2)}}w_{k_{(2)}m_{(2)}r_{(2)}} \\
				&\quad\quad\quad\quad+ 
				\frac{\theta_{3}-1}{6}q_{j_{(1)}k_{(2)}}\sum_{r_{(2)}=1}^{N_{(2)}}w_{k_{(2)}r_{(2)}m_{(2)}}+\frac{\theta_{3}-1}{6}\sum_{n_{(2)}=1}^{N_{(2)}}q_{j_{(1)}n_{(2)}}w_{n_{(2)}k_{(2)}m_{(2)}} \\
				&\quad\quad\quad\quad-
				\sum_{u_{(1)}=1}^{N_{(1)}}p_{i_{(1)}u_{(1)}}^{(1)}\biggl(\frac{\theta_{2}-1}{2}q_{u_{(1)}k_{(2)}}w_{k_{(2)}m_{(2)}} \\
				&\quad\quad\quad\quad+ 
				\frac{\theta_{3}-1}{6}q_{u_{(1)}k_{(2)}}\sum_{r_{(2)}=1}^{N_{(2)}}w_{k_{(2)}m_{(2)}r_{(2)}}+\frac{\theta_{3}-1}{6}q_{u_{(1)}k_{(2)}}\sum_{r_{(2)}=1}^{N_{(2)}}w_{k_{(2)}r_{(2)}m_{(2)}} \\
				&\quad\quad\quad\quad+ 
				\frac{\theta_{3}-1}{6}\sum_{n_{(2)}=1}^{N_{(2)}}q_{u_{(1)}n_{(2)}}w_{n_{(2)}k_{(2)}m_{(2)}}\biggr) \biggr) \biggr),\\
			\end{aligned}
		\end{equation}
		\begin{equation}\label{eq:example2}
			\begin{aligned}
				&M_{k_{(2)}m_{(2)}n_{(2)}}^{i_{(1)}j_{(1)}} =
				(1-\alpha){p_{i_{(1)}j_{(1)}}^{(1)}}\biggl(b\biggl(\frac{{(\theta_{3}-1)}^{2}}{6}q_{j_{(1)}k_{(2)}}w_{k_{(2)}m_{(2)}n_{(2)}} \\
				&\quad\quad\quad\quad-
				\sum_{u_{(1)}=1}^{N_{(1)}}p_{i_{(1)}u_{(1)}}^{(1)}\biggl(\frac{{(\theta_{3}-1)}^{2}}{6}q_{u_{(1)}k_{(2)}}w_{k_{(2)}m_{(2)}n_{(2)}}\biggr) \biggr) \biggr),\\
			\end{aligned}
		\end{equation}
		For simplicity, we let $N_{(1)}=N_{(2)}=N$,
		where 
		\begin{equation}\label{eq:example}
			\psi_{i_{(1)}j_{(1)}} = \frac{N}{2}(\hat{\eta}_{(1)} - \eta_{i_{(1)}j_{(1)}}) + \frac{1}{2}\sum_{k_{(1)}=1}^{N} \left(p_{i_{(1)}k_{(1)}}^{(1)}\psi_{k_{(1)}j_{(1)}} + p_{j_{(1)}k_{(1)}}^{(1)}\psi_{i_{(1)}k_{(1)}}\right),
		\end{equation}
		\begin{equation}\label{eq:example}
			\begin{aligned}
				&\nu_{i_{(1)}j_{(1)}h_{(1)}} = 
				\frac{N}{3}(\hat{\eta}_{(1)} - \eta_{i_{(1)}j_{(1)}h_{(1)}}) + \frac{1}{3}\sum_{k_{(1)}=1}^{N}\biggr(p_{i_{(1)}k_{(1)}}^{(1)}\nu_{k_{(1)}j_{(1)}h_{(1)}}+p_{j_{(1)}k_{(1)}}^{(1)}\nu_{i_{(1)}k_{(1)}h_{(1)}} \\
				&\quad\quad\quad\quad+
				p_{h_{(1)}k_{(1)}}^{(1)}\nu_{i_{(1)}j_{(1)}k_{(1)}} \biggr),
			\end{aligned}
		\end{equation}
		\begin{equation}\label{eq:example}
			\begin{aligned}
				&\kappa_{i_{(1)}j_{(1)}h_{(1)}l_{(1)}} = 
				\frac{N}{4}(\hat{\eta}_{(1)} - \eta_{i_{(1)}j_{(1)}h_{(1)}l_{(1)}}) + \frac{1}{4}\sum_{k_{(1)}=1}^{N}\biggr(p_{i_{(1)}k_{(1)}}^{(1)}\kappa_{k_{(1)}j_{(1)}h_{(1)}l_{(1)}} \\
				&\quad\quad\quad\quad+
				p_{j_{(1)}k_{(1)}}^{(1)}\kappa_{i_{(1)}k_{(1)}h_{(1)}l_{(1)}}+p_{h_{(1)}k_{(1)}}^{(1)}\kappa_{i_{(1)}j_{(1)}k_{(1)}l_{(1)}}+p_{l_{(1)}k_{(1)}}^{(1)}\kappa_{i_{(1)}j_{(1)}h_{(1)}k_{(1)}}\biggr),
			\end{aligned}
		\end{equation}
		\begin{equation}\label{eq:example}
			\begin{aligned}
				&\phi_{i_{(1)}j_{(2)}} = \frac{N^{2}}{2N-1}((\hat{\eta}_{(1)}\hat{\eta}_{(2)}) - \eta_{i_{(1)}j_{(2)}}) \\
				&\quad\quad\quad\quad+
				\frac{1}{2N-1}\sum_{k_{(1)}=1}^{N}\sum_{k_{(2)}=1}^{N} p_{i_{(1)}k_{(1)}}^{(1)} p_{j_{(2)}k_{(2)}}^{(1)} \phi_{k_{(1)}k_{(2)}} \\
				&\quad\quad\quad\quad+
				\frac{N-1}{2N-1}\sum_{k_{(1)}=1}^{N}p_{i_{(1)}k_{(1)}}^{(1)}\phi_{k_{(1)}j_{(2)}}+\frac{N-1}{2N-1}\sum_{k_{(2)}=1}^{N}p_{j_{(2)}k_{(2)}}^{(1)}\phi_{i_{(1)}k_{(2)}},
			\end{aligned}
		\end{equation}
		\begin{equation}\label{eq:example}
			\begin{aligned}
				&\sigma_{i_{(1)}j_{(2)}h_{(2)}} = \frac{N^{2}}{3N-2}(\hat{\eta}_{(1)}\hat{\eta}_{(2)} - \eta_{i_{(1)}j_{(2)}h_{(2)}}) \\
				&\quad\quad\quad\quad+ \frac{1}{3N-2}\sum_{k_{(1)}=1}^{N}\sum_{k_{(2)}=1}^{N} p_{i_{(1)}k_{(1)}}^{(1)} p_{j_{(2)}k_{(2)}}^{(1)} \sigma_{k_{(1)}k_{(2)}h_{(2)}} \\
				&\quad\quad\quad\quad+
				\frac{1}{3N-2}\sum_{k_{(1)}=1}^{N}\sum_{k_{(2)}=1}^{N} p_{i_{(1)}k_{(1)}}^{(1)} p_{h_{(2)}k_{(2)}}^{(1)} \sigma_{k_{(1)}j_{(2)}k_{(2)}} \\
				&\quad\quad\quad\quad+ \frac{N-2}{3N-2}\sum_{k_{(1)}=1}^{N}p_{i_{(1)}k_{(1)}}^{(1)}\sigma_{k_{(1)}j_{(2)}h_{(2)}}+\frac{N-1}{3N-2}\sum_{k_{(2)}=1}^{N}p_{j_{(2)}k_{(2)}}^{(1)}\sigma_{i_{(1)}k_{(2)}h_{(2)}} \\
				&\quad\quad\quad\quad+ \frac{N-1}{3N-2}\sum_{k_{(2)}=1}^{N}p_{h_{(2)}k_{(2)}}^{(1)}\sigma_{i_{(1)}j_{(2)}k_{(2)}},
			\end{aligned}
		\end{equation}
		\begin{equation}\label{eq:example}
			\begin{aligned}
				&\upsilon_{i_{(1)}j_{(2)}h_{(2)}r_{(2)}} = \frac{N^{2}}{4N-3}(\hat{\eta}_{(1)}\hat{\eta}_{(2)} - \eta_{i_{(1)}j_{(2)}h_{(2)}r_{(2)}}) \\
				&\quad\quad\quad\quad+
				\frac{1}{4N-3}\sum_{k_{(1)}=1}^{N}\sum_{k_{(2)}=1}^{N} p_{i_{(1)}k_{(1)}}^{(1)}p_{j_{(2)}k_{(2)}}^{(1)}\upsilon_{k_{(1)}k_{(2)}h_{(2)}r_{(2)}} \\
				&\quad\quad\quad\quad+
				\frac{1}{4N-3}\sum_{k_{(1)}=1}^{N}\sum_{k_{(2)}=1}^{N} p_{i_{(1)}k_{(1)}}^{(1)} p_{h_{(2)}k_{(2)}}^{(1)}\upsilon_{k_{(1)}j_{(2)}k_{(2)}r_{(2)}} \\
				&\quad\quad\quad\quad+
				\frac{1}{4N-3}\sum_{k_{(1)}=1}^{N}\sum_{k_{(2)}=1}^{N} p_{i_{(1)}k_{(1)}}^{(1)} p_{r_{(2)}k_{(2)}}^{(1)} \upsilon_{k_{(1)}j_{(2)}h_{(2)}k_{(2)}} \\
				&\quad\quad\quad\quad+
				\frac{N-3}{4N-3}\sum_{k_{(1)}=1}^{N}p_{i_{(1)}k_{(1)}}^{(1)}\upsilon_{k_{(1)}j_{(2)}h_{(2)}r_{(2)}}+\frac{N-1}{4N-3}\sum_{k_{(2)}=1}^{N}p_{j_{(2)}k_{(2)}}^{(1)}\upsilon_{i_{(1)}k_{(2)}h_{(2)}r_{(2)}} \\
				&\quad\quad\quad\quad+
				\frac{N-1}{4N-3}\sum_{k_{(2)}=1}^{N}p_{h_{(2)}k_{(2)}}^{(1)}\upsilon_{i_{(1)}j_{(2)}k_{(2)}r_{(2)}}+\frac{N-1}{4N-3}\sum_{k_{(2)}=1}^{N}p_{r_{(2)}k_{(2)}}^{(1)}\upsilon_{i_{(1)}j_{(2)}h_{(2)}k_{(2)}},
			\end{aligned}
		\end{equation}
	\end{corollary}
	The above expressions for $\psi_{i_{(1)}j_{(1)}}, \nu_{i_{(1)}j_{(1)}h_{(1)}}, \kappa_{i_{(1)}j_{(1)}h_{(1)}l_{(1)}}, \phi_{i_{(1)}j_{(2)}},\sigma_{i_{(1)}j_{(2)}h_{(2)}},\upsilon_{i_{(1)}j_{(2)}h_{(2)}r_{(2)}},\zeta_{i_{(2)}j_{(1)}},\xi_{i_{(2)}j_{(2)}}$ are only satisfied when the variables such as $i_{(1)},j_{(2)},h_{(1)}$ are not equal to each other. Due to the complexity of the equations, when there is an equal relationship between variables such as $i_{(1)},j_{(2)},h_{(1)}$, the corresponding expressions can be easily obtained, so they will not be repeated.
	When we calculate the critical value $(\frac{b}{c})^{*}$, to make the results more convincing, we can obtain expressions for some quantities\citep{mcavoy2021fixation},
	\begin{equation}\label{eq:example}
		\begin{aligned}
			&\upsilon_{\textbf{I}}^{(1)}= E_{{\lambda}_{D}}^{\circ}[{\rho^{\circ}_{{\textbf{C}}_{(1)} \times {\textbf{C}}_{(2)}}(\bm{\eta})+\rho^{\circ}_{{\textbf{D}}_{(1)} \times {\textbf{C}}_{(2)}}(\bm{\eta})}](E_{{\lambda}}^{\circ}[\hat{\eta}_{(1)}\hat{\eta}_{(2)}]-E_{{\lambda}}^{\circ}[\bm{\eta}_{\textbf{I}}]) \\
			&\quad\quad\quad\quad+
			E_{{\lambda}}^{\circ}[{\rho^{\circ}_{{\textbf{C}}_{(1)} \times {\textbf{C}}_{(2)}}(\bm{\eta})+\rho^{\circ}_{{\textbf{C}}_{(1)} \times {\textbf{D}}_{(2)}}(\bm{\eta})}](E_{{\lambda}_{D}}^{\circ}[\hat{\eta}_{(1)}\hat{\eta}_{(2)}]-E_{{\lambda}_{D}}^{\circ}[\bm{\eta}_{\textbf{I}}]) \\
			&\quad\quad\quad\quad+
			\sum_{(Q_{(1)},f_{(1)}),(Q_{(2)},f_{(2)})}p_{(Q_{(1)},f_{(1)})}^{\circ}p_{(Q_{(2)},f_{(2)})}^{\circ}\upsilon_{{\tilde{f}_{(1)}}\times {\tilde{f}_{(2)}}(\textbf{I})}^{(1)},
		\end{aligned}
	\end{equation}
	the expressions for $\sigma_{\textbf{I}}^{(1)},\phi_{\textbf{I}}^{(1)},\zeta_{\textbf{I}}^{(1)}$ are similar,
	\begin{equation}\label{eq:example}
		\begin{aligned}
			&\kappa_{\textbf{I}}^{(1)}= E_{{\lambda}_{D}}^{\circ}[{\rho^{\circ}_{{\textbf{C}}_{(1)} \times {\textbf{C}}_{(2)}}(\bm{\eta})+\rho^{\circ}_{{\textbf{D}}_{(1)} \times {\textbf{C}}_{(2)}}(\bm{\eta})}](E_{{\lambda}}^{\circ}[\hat{\eta}_{(1)}]-E_{{\lambda}}^{\circ}[\bm{\eta}_{\textbf{I}}]) \\
			&\quad\quad\quad\quad+
			E_{{\lambda}}^{\circ}[{\rho^{\circ}_{{\textbf{C}}_{(1)} \times {\textbf{C}}_{(2)}}(\bm{\eta})+\rho^{\circ}_{{\textbf{C}}_{(1)} \times {\textbf{D}}_{(2)}}(\bm{\eta})}](E_{{\lambda}_{D}}^{\circ}[\hat{\eta}_{(1)}]-E_{{\lambda}_{D}}^{\circ}[\bm{\eta}_{\textbf{I}}]) \\
			&\quad\quad\quad\quad+
			\sum_{(Q_{(1)},f_{(1)}),(Q_{(2)},f_{(2)})}p_{(Q_{(1)},f_{(1)})}^{\circ}p_{(Q_{(2)},f_{(2)})}^{\circ}\kappa_{{\tilde{f}_{(1)}}\times {\tilde{f}_{(2)}}(\textbf{I})}^{(1)},
		\end{aligned}
	\end{equation}
	the expressions for$\nu_{\textbf{I}}^{(1)},\psi_{\textbf{I}}^{(1)},\xi_{\textbf{I}}^{(1)}$ are similar.
	we could use new expressions about $\kappa_{\textbf{I}}^{(1)},\upsilon_{\textbf{I}}^{(1)},\nu_{\textbf{I}}^{(1)},\psi_{\textbf{I}}^{(1)},\xi_{\textbf{I}}^{(1)},\sigma_{\textbf{I}}^{(1)},\phi_{\textbf{I}}^{(1)},\zeta_{\textbf{I}}^{(1)}$ to judge based on condition\citep{mcavoy2021fixation}
	\begin{equation}\label{eq:example}
		\frac{d}{d\delta} \bigg|_{\delta=0}\frac{E_{{\lambda}}[{\rho^{\circ}_{{\textbf{C}}_{(1)} \times {\textbf{C}}_{(2)}}(\bm{\eta})+\rho^{\circ}_{{\textbf{C}}_{(1)} \times {\textbf{D}}_{(2)}}(\bm{\eta})}]}{E_{{\lambda}}[{\rho^{\circ}_{{\textbf{C}}_{(1)} \times {\textbf{C}}_{(2)}}(\bm{\eta})+\rho^{\circ}_{{\textbf{C}}_{(1)} \times {\textbf{D}}_{(2)}}(\bm{\eta})}]+E_{{\lambda}_{D}}[{\rho^{\circ}_{{\textbf{C}}_{(1)} \times {\textbf{C}}_{(2)}}(\bm{\eta})+\rho^{\circ}_{{\textbf{D}}_{(1)} \times {\textbf{C}}_{(2)}}(\bm{\eta})}]} > 0
	\end{equation}
	does the network structure support cooperative behavior under weak selection in two-order and higher-order game.
	Due to the complexity of the formula, we will not elaborate on the result in $l_{(2)}$ network.
	\section{Conclusion}
	In this article, we obtain the calculation formula for the fixation probability of two-order and higher-order game in a two-layer network under weak selection through complex theoretical analysis,as well as the optimal bias coefficient of utility function $\alpha$ for two-order game in some two-layer networks.We also conclude that when the bias coefficient of utility function $\alpha$ satisfies certain conditions,in some two-layer networks promote cooperation more than in some single-layer networks.This provides a reliable theoretical basis for subsequent research on the evolution of cooperative behavior in the network.In our research,we mainly consider the fixed probability under weak selection,without considering any selection coefficient, which is a good extension direction.In addition,we have introduced low mutation without considering the theoretical results under any mutation or no mutation,which is the future research direction.

	\bibliographystyle{elsarticle-harv}
	
	\bibliography{refer}

\end{document}